\documentclass[conference,compsoc]{IEEEtran}
\usepackage{float}
\usepackage{algorithm}
\usepackage{algorithmic}
\usepackage{amsmath,amssymb,amsfonts}
\usepackage{graphicx}
\usepackage{textcomp}
\usepackage{colortbl}
\usepackage{flushend}
\usepackage{adjustbox}
\usepackage[utf8]{inputenc}
\usepackage{pgfplots}
\DeclareUnicodeCharacter{2212}{−}
\usepgfplotslibrary{groupplots,dateplot}
\usetikzlibrary{patterns,shapes.arrows}
\pgfplotsset{compat=newest}
\usepackage{caption}
\usepackage{tabularx}
\usepackage{import}
\usepackage{listings}
\usepackage{multirow}
\usepackage{caption}
\usepackage{subcaption}
\usepackage{hyperref}
\usepackage{circuitikz}
\usepackage{pgf}
\usepackage{pgfplots}
\usepackage{xcolor}
\usepackage{bm}
\usepackage{pgfplotstable}
\pgfplotsset{compat=1.8}
\usepackage{adjustbox}
\usepackage{changepage}
\usepackage{dblfloatfix}
\usepackage{tcolorbox}
\usepackage{listing}
\usepackage{xpatch}
\usepackage{booktabs}
\usepackage[utf8]{inputenc}
\usepackage{enumitem}

\definecolor{bblue}{HTML}{235999}
\definecolor{rred}{HTML}{A35966}
\definecolor{ggreen}{HTML}{417505}
\definecolor{ppurple}{HTML}{7859A3}
\definecolor{ggray}{HTML}{4A4A4A}
\definecolor{llightblue}{HTML}{66CCFF}

\makeatletter

\tikzstyle{chart}=[
    legend label/.style={font={\scriptsize},anchor=west,align=left},
    legend box/.style={rectangle, draw, minimum size=5pt},
    axis/.style={black,semithick,->},
    axis label/.style={anchor=east,font={\tiny}},
]

\tikzstyle{bar chart}=[
    chart,
    bar width/.code={
        \pgfmathparse{##1/2}
        \global\let\bar@w\pgfmathresult
    },
    bar/.style={very thick, draw=white},
    bar label/.style={font={\bf\small},anchor=north},
    bar value/.style={font={\footnotesize}},
    bar width=.75,
]

\tikzstyle{pie chart}=[
    chart,
    slice/.style={line cap=round, line join=round, very thick,draw=white},
    pie title/.style={font={\bf},shift={(0cm,-0.2cm)}},
    slice type/.style 2 args={
        ##1/.style={fill=##2},
        values of ##1/.style={}
    }
]

\pgfdeclarelayer{background}
\pgfdeclarelayer{foreground}
\pgfsetlayers{background,main,foreground}

\newcommand{\pie}[3][]{
    \begin{scope}[#1]
    \pgfmathsetmacro{\curA}{90}
    \pgfmathsetmacro{\r}{1}
    \def\c{(0,0)}
    \node[pie title] at (90:1.3) {#2};
    \foreach \v/\s in{#3}{
        \pgfmathsetmacro{\deltaA}{\v/100*360}
        \pgfmathsetmacro{\nextA}{\curA + \deltaA}
        \pgfmathsetmacro{\midA}{(\curA+\nextA)/2}

        \path[slice,\s] \c
            -- +(\curA:\r)
            arc (\curA:\nextA:\r)
            -- cycle;
        \pgfmathsetmacro{\d}{max((\deltaA * -(.5/50) + 1) , .5)}

        \begin{pgfonlayer}{foreground}
        \path \c -- node[pos=\d,pie values,values of \s]{$\v\%$} +(\midA:\r);
        \end{pgfonlayer}

        \global\let\curA\nextA
    }
    \end{scope}
}

\newcommand{\legend}[2][]{
    \begin{scope}[#1]
    \path
        \foreach \n/\s in {#2}
            {
                  ++(0,-10pt) node[\s,legend box] {} +(2pt,0) node[legend label] {\n}
            }
    ;
    \end{scope}
}

\newcommand{\INPUT}{\item[\textbf{Input:}]}
\newcommand{\OUTPUT}{\item[\textbf{Output:}]}

\ifCLASSOPTIONcompsoc
  \usepackage[nocompress]{cite}
\else
  \usepackage{cite}
\fi

\ifCLASSINFOpdf
\else
\fi

\newcommand\xqed[1]{%
  \leavevmode\unskip\penalty9999 \hbox{}\nobreak\hfill
  \quad\hbox{#1}}
\newcommand\exampleEnd{\xqed{$\blacksquare$}}

\newcounter{comment}

\newcounter{response}

\begin{document}

\title{SysFuSS: System-Level Firmware Fuzzing with Selective Symbolic Execution}

\vspace{-0.2in}
\author{
{\rm Dakshina Tharindu}\\
University of Florida
\and
{\rm Aruna Jayasena}\\
University of Florida
\and
{\rm Prabhat Mishra}\\
University of Florida
} 

\maketitle

\begin{abstract}
Firmware serves as the critical interface between hardware and software in computing systems, making any bugs or vulnerabilities particularly dangerous as they can cause catastrophic system failures. While fuzzing is a promising approach for identifying design flaws and security vulnerabilities, traditional fuzzers are ineffective at detecting firmware vulnerabilities. For example, existing fuzzers focus on user-level fuzzing, which is not suitable for detecting kernel-level vulnerabilities. Existing fuzzers also face a coverage plateau problem when dealing with complex interactions between firmware and hardware. In this paper, we present an efficient firmware verification framework, SysFuSS, that integrates system-level fuzzing with selective symbolic execution. Our approach leverages system-level emulation for initial fuzzing, and automatically transitions to symbolic execution when coverage reaches a plateau. This strategy enables us to generate targeted test cases that can trigger previously unexplored regions in firmware designs. We have evaluated SysFuSS on real-world embedded firmware, including OpenSSL, WolfBoot, WolfMQTT, HTSlib, MXML, and libIEC. Experimental evaluation demonstrates that SysFuSS significantly outperforms state-of-the-art fuzzers in terms of both branch coverage and detection of firmware vulnerabilities. 
Specifically, SysFuSS can detect 118 known vulnerabilities while state-of-the-art can cover only 13 of them. Moreover, SysFuSS takes significantly less time (up to 3.3X, 1.7X on average) to activate these vulnerabilities.   




\end{abstract}


%
\IEEEpeerreviewmaketitle

\section{Introduction} \label{sec:introduction}

Firmware is the invisible foundation of computing devices, from tiny edge devices to enterprise servers. It operates at the most privileged level, bridging hardware and software to provide low-level control of system components. This critical role, however, makes firmware a prime target for attackers and a challenging frontier for defenders. Firmware often consists of opaque, proprietary code running with deep system access, which complicates security analysis. The explosive growth of the Internet of Things (IoT) has amplified this issue. IoT devices introduce a massive attack surface in homes, industries, and critical infrastructure. For example, researchers have identified over 540,000 publicly accessible IoT devices in 144 countries configured with factory default root passwords and vulnerable to known exploits~\cite{10.1145/1920261.1920276}.  
These trends highlight the urgent need for firmware security validation.


          
         
          
         

\begin{figure}[htbp]
\centering
            \hspace{1cm}
            \begin{subfigure}{.5\linewidth}
          \centering
          \tikzset{every picture/.style={line width=0.75pt}} 

\begin{tikzpicture}[x=0.75pt,y=0.75pt,yscale=-1,xscale=1]

\draw  [fill={rgb, 255:red, 255; green, 243; blue, 217 }  ,fill opacity=1 ] (60,135.13) .. controls (60,129.54) and (64.54,125) .. (70.13,125) -- (129.87,125) .. controls (135.46,125) and (140,129.54) .. (140,135.13) -- (140,244.87) .. controls (140,250.46) and (135.46,255) .. (129.87,255) -- (70.13,255) .. controls (64.54,255) and (60,250.46) .. (60,244.87) -- cycle ;
\draw  [fill={rgb, 255:red, 255; green, 243; blue, 217 }  ,fill opacity=1 ] (260,135.13) .. controls (260,129.54) and (264.54,125) .. (270.13,125) -- (329.87,125) .. controls (335.46,125) and (340,129.54) .. (340,135.13) -- (340,244.87) .. controls (340,250.46) and (335.46,255) .. (329.87,255) -- (270.13,255) .. controls (264.54,255) and (260,250.46) .. (260,244.87) -- cycle ;
\draw  [fill={rgb, 255:red, 255; green, 221; blue, 166 }  ,fill opacity=1 ] (230,189.61) -- (170.2,190) -- (170,148.89) .. controls (207.38,148.65) and (199.84,133.88) .. (229.78,143.28) -- cycle ;
\draw  [fill={rgb, 255:red, 255; green, 178; blue, 105 }  ,fill opacity=1 ] (170.2,190) -- (230,190) -- (230,231.25) .. controls (192.62,231.25) and (200.1,246.13) .. (170.2,236.5) -- cycle ;
\draw  [fill={rgb, 255:red, 255; green, 221; blue, 166 }  ,fill opacity=1 ] (130,209.61) -- (70.19,210) -- (70,168.89) .. controls (107.38,168.65) and (99.83,153.87) .. (129.78,163.27) -- cycle ;
\draw  [fill={rgb, 255:red, 255; green, 221; blue, 166 }  ,fill opacity=1 ] (330,184.61) -- (270.19,185) -- (270,143.89) .. controls (307.38,143.65) and (299.83,128.87) .. (329.78,138.27) -- cycle ;
\draw  [fill={rgb, 255:red, 255; green, 178; blue, 105 }  ,fill opacity=1 ] (270.2,185) -- (330,185) -- (330,226.25) .. controls (292.62,226.25) and (300.1,241.13) .. (270.2,231.5) -- cycle ;
\draw  [fill={rgb, 255:red, 222; green, 222; blue, 255 }  ,fill opacity=1 ] (60,69) .. controls (60,64.03) and (64.03,60) .. (69,60) -- (131,60) .. controls (135.97,60) and (140,64.03) .. (140,69) -- (140,96) .. controls (140,100.97) and (135.97,105) .. (131,105) -- (69,105) .. controls (64.03,105) and (60,100.97) .. (60,96) -- cycle ;
\draw    (90,105) -- (90,122) ;
\draw [shift={(90,125)}, rotate = 270] [fill={rgb, 255:red, 0; green, 0; blue, 0 }  ][line width=0.08]  [draw opacity=0] (10.72,-5.15) -- (0,0) -- (10.72,5.15) -- (7.12,0) -- cycle    ;
\draw    (110,125) -- (110,108) ;
\draw [shift={(110,105)}, rotate = 90] [fill={rgb, 255:red, 0; green, 0; blue, 0 }  ][line width=0.08]  [draw opacity=0] (10.72,-5.15) -- (0,0) -- (10.72,5.15) -- (7.12,0) -- cycle    ;
\draw    (290,105) -- (290,122) ;
\draw [shift={(290,125)}, rotate = 270] [fill={rgb, 255:red, 0; green, 0; blue, 0 }  ][line width=0.08]  [draw opacity=0] (10.72,-5.15) -- (0,0) -- (10.72,5.15) -- (7.12,0) -- cycle    ;
\draw    (310,125) -- (310,108) ;
\draw [shift={(310,105)}, rotate = 90] [fill={rgb, 255:red, 0; green, 0; blue, 0 }  ][line width=0.08]  [draw opacity=0] (10.72,-5.15) -- (0,0) -- (10.72,5.15) -- (7.12,0) -- cycle    ;
\draw [line width=2.25]    (170,170) -- (145,170) ;
\draw [shift={(140,170)}, rotate = 360] [fill={rgb, 255:red, 0; green, 0; blue, 0 }  ][line width=0.08]  [draw opacity=0] (8.57,-4.12) -- (0,0) -- (8.57,4.12) -- cycle    ;
\draw [line width=2.25]    (240,185) -- (255,185) ;
\draw [shift={(260,185)}, rotate = 180] [fill={rgb, 255:red, 0; green, 0; blue, 0 }  ][line width=0.08]  [draw opacity=0] (8.57,-4.12) -- (0,0) -- (8.57,4.12) -- cycle    ;
\draw [line width=2.25]    (230,165) -- (240,165) -- (240,185) ;
\draw [line width=2.25]    (240,185) -- (240,210) -- (230,210) ;
\draw  [fill={rgb, 255:red, 222; green, 222; blue, 255 }  ,fill opacity=1 ] (260,69) .. controls (260,64.03) and (264.03,60) .. (269,60) -- (331,60) .. controls (335.97,60) and (340,64.03) .. (340,69) -- (340,96) .. controls (340,100.97) and (335.97,105) .. (331,105) -- (269,105) .. controls (264.03,105) and (260,100.97) .. (260,96) -- cycle ;
\draw  [draw opacity=0] (40,145) -- (55,145) -- (55,185) -- (40,185) -- cycle ;

\draw (100,243.86) node  [font=\small] [align=left] {\begin{minipage}[lt]{54.4pt}\setlength\topsep{0pt}
\begin{center}
\textbf{Emulator}
\end{center}

\end{minipage}};
\draw (300,243.44) node  [font=\small] [align=left] {\begin{minipage}[lt]{54.4pt}\setlength\topsep{0pt}
\begin{center}
\textbf{Emulator}
\end{center}

\end{minipage}};
\draw (200,167.3) node  [font=\small] [align=left] {\begin{minipage}[lt]{40.8pt}\setlength\topsep{0pt}
\begin{center}
User level code
\end{center}

\end{minipage}};
\draw (200,212.5) node  [font=\small] [align=left] {\begin{minipage}[lt]{40.8pt}\setlength\topsep{0pt}
\begin{center}
Kernel code
\end{center}

\end{minipage}};
\draw (100,187.5) node  [font=\small] [align=left] {\begin{minipage}[lt]{40.8pt}\setlength\topsep{0pt}
\begin{center}
User level code
\end{center}

\end{minipage}};
\draw (300,162.5) node  [font=\small] [align=left] {\begin{minipage}[lt]{40.8pt}\setlength\topsep{0pt}
\begin{center}
User level code
\end{center}

\end{minipage}};
\draw (300,207.5) node  [font=\small] [align=left] {\begin{minipage}[lt]{40.8pt}\setlength\topsep{0pt}
\begin{center}
Kernel code
\end{center}

\end{minipage}};
\draw (100,82.5) node  [font=\small] [align=left] {\begin{minipage}[lt]{54.4pt}\setlength\topsep{0pt}
\begin{center}
\textbf{Fuzzer}
\end{center}

\end{minipage}};
\draw (200,125) node   [align=left] {\begin{minipage}[lt]{54.4pt}\setlength\topsep{0pt}
\begin{center}
\textit{Firmware Binary}
\end{center}

\end{minipage}};
\draw (300,82.5) node  [font=\small] [align=left] {\begin{minipage}[lt]{54.4pt}\setlength\topsep{0pt}
\begin{center}
\textbf{Fuzzer}
\end{center}

\end{minipage}};

\end{tikzpicture}
         
          \vspace{-0.1in}
          \caption{State-of-the-art}
          \label{fig:existing}
        \end{subfigure}%
        \begin{subfigure}{.5\linewidth}
          \centering

          \vspace{-0.05in}
          \caption{Proposed (SysFuSS)}
          \label{fig:proposed}
        \end{subfigure}

\caption{State-of-the-art fuzzers can only handle user-level programs and cannot handle the system-level interactions. In contrast, our proposed approach can handle both system-level and user-level programs.}
\label{fig:intro-limits}
\end{figure}
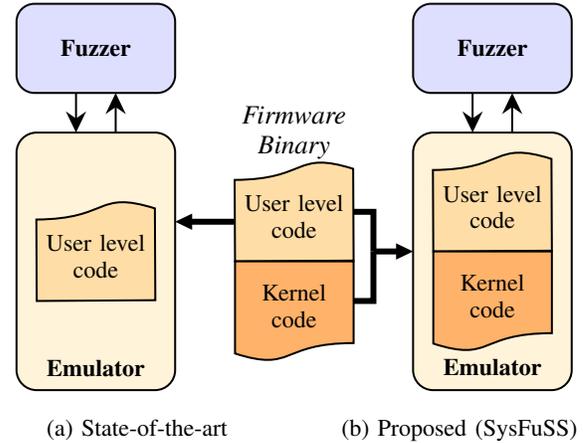

Fuzzing is widely used for uncovering software vulnerabilities. Coverage-guided fuzzers, such as AFL~\cite{afl} and AFL++,~\cite{fioraldi2020afl++} can automatically generate effective test patterns to expose subtle bugs in software programs. Their success has inspired extensions of fuzzing beyond traditional user-space software. For example, AFLNet~\cite{pham2020aflnet} adapts coverage-guided fuzzing to stateful network protocol implementations by incorporating protocol-aware mutations and message sequencing. Recent fuzzers, such as AFLFast~\cite{bohme2016coverage} and EM-Fuzz~\cite{gao2020fuzz}, have proposed firmware-oriented fuzzing, as illustrated by Figure~\ref{fig:existing}.

While state-of-the-art firmware fuzzing techniques have shown promise for user-level functionalities, they face significant limitations when applied to complete firmware binaries that include kernel-level functionality. In particular, existing implementations often lack the infrastructure for full system-level emulation and do not incorporate effective state-space management strategies to handle the vast execution space arising from complex hardware–firmware interactions. As a result, these approaches struggle to exercise deeply embedded code paths, and the progress of the fuzzer eventually stagnates, manifesting as coverage plateaus where hard-to-trigger corner cases remain unexplored.

\subsection{Contributions}
\vspace{-0.1in}
To address these challenges, we propose SysFuSS, a hybrid framework that combines system-level fuzzing with selective symbolic execution for comprehensive firmware analysis, as illustrated by Figure~\ref{fig:proposed}. SysFuSS leverages the complementary strengths of both techniques: fuzzing rapidly explores broad execution paths, while symbolic execution systematically solves complex constraints to reach deep, hard-to-trigger states. SysFuSS integrates these methods in a feedback-driven loop, executing full firmware in emulation to collect coverage data, detecting when progress stalls, and selectively invoking symbolic reasoning to generate new input vectors. These input vectors are then fed back to the fuzzer, enabling SysFuSS to overcome coverage plateaus and achieve deeper exploration without the continuous overhead of symbolic analysis. Specifically, this work makes the following key contributions:
\begin{enumerate}[topsep=5pt, itemsep=3pt, parsep=3pt, leftmargin=*]
\item \textbf{System-level emulation:} A QEMU-based execution environment that treats the entire firmware binary, including kernel and driver code, as a single analyzable unit.
\item \textbf{Memory-aware fuzzing:} Lightweight instrumentation for detecting illegal heap and stack accesses within the emulated firmware binary.
\item \textbf{Coverage plateau detection:} A runtime mechanism that monitors coverage trends and triggers symbolic execution only when the fuzzer reaches a coverage plateau.
\item \textbf{Guided symbolic reasoning:} A strategy to leverage the fuzzing trajectory to focus symbolic analysis of hard-to-reach program regions to avoid state-space explosion.
\item \textbf{Comprehensive evaluation:} An empirical study on six real-world firmware benchmarks demonstrating significant improvement in branch coverage and detection of vulnerabilities over state-of-the-art approaches.
\end{enumerate}
\vspace{1mm}

SysFuSS advances the state-of-the-art in firmware security analysis by introducing a practical hybrid framework that combines the high-throughput exploration of system-level fuzzing with the precise path-solving capability of symbolic execution. This enables SysFuSS to uncover deep, previously unreachable vulnerabilities in embedded firmware implementations.
The remainder of this paper is organized as follows. Section~\ref{sec:background} presents the necessary background and surveys related efforts. Section~\ref{sec:methodology} describes the design and implementation details of the proposed SysFuSS framework. Section~\ref{sec:experiemnts} presents the experimental results. Finally, Section~\ref{sec:conclusion} concludes the paper.

\section{Background}\label{sec:background}

In this section, we first motivate the need for system-level fuzzing by reviewing recent trends in firmware vulnerabilities. We then provide an overview of coverage-guided greybox fuzzing, followed by a discussion of user-level and system-level emulation. Finally, we survey existing research efforts in firmware fuzzing to position our work within the current landscape.

\subsection{Survey of Firmware Vulnerabilities}

Firmware serves as the foundational software layer that directly interfaces with hardware components. Due to its privileged position in the system stack, any vulnerabilities present in firmware can be particularly dangerous. Exploits at this level have the potential to bypass higher-level security mechanisms, such as access controls, secure communication protocols, and privilege management. Once compromised, firmware can serve as a stealthy foothold for attackers, often persisting undetected even through system reboots or operating system reinstalls.

There are repositories, such as MITRE Common Vulnerabilities and Exposures (CVE)~\cite{cveorg} and NIST National Vulnerability Database (NVD)~\cite{nistnvd}, for centralized listings of disclosed vulnerabilities. However, it is a major challenge to track and analyze the firmware-specific vulnerabilities. This difficulty arises due to practical constraints in real-world systems, including the lack of publicly available source code, restricted observability into the firmware's internal logic, and the intricate dependencies between firmware and hardware peripherals. As a result, many vulnerabilities remain undiscovered until after deployment or are identified through third-party security audits long after a device has reached end users.
To better understand the current threat landscape, we conducted a systematic survey of firmware-related CVEs reported from 2012 to 2025. Our data collection, sourced from the CVE.org database~\cite{cveorg}, identified a total of 6,612 CVEs associated with firmware across this 13-year span. To make the analysis actionable, we categorized these vulnerabilities into six representative classes: use-after-free, heap-based buffer overflows, stack-based buffer overflows, out-of-bound writes, improper input validation, and OS command injection. These categories represent the dominant types of vulnerabilities frequently observed in firmware across different vendors and platforms.

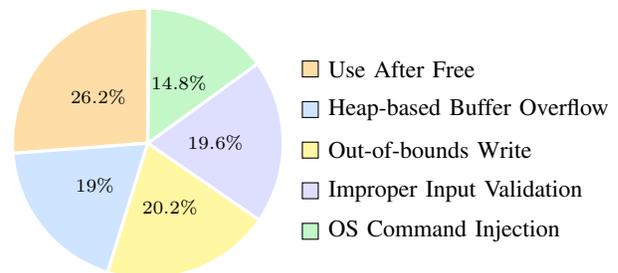
\begin{figure}[htp]
\centering
\small
\vspace{-0.2in}
\definecolor{lightred}{RGB}{255,221,166}
\definecolor{lightblue}{RGB}{207,228,255}
\definecolor{lightgray}{RGB}{195,247,200}
\definecolor{lightpurple}{RGB}{222,222,255}
\definecolor{lightgreen}{RGB}{255,247,161}

\begin{tikzpicture}
[
    pie chart,
    slice type={cwe416}{lightred},
    slice type={cwe122}{lightblue},
    slice type={cwe787}{lightgreen},
    slice type={cwe20}{lightpurple},
    slice type={cwe78}{lightgray},
    pie values/.style={font={\color{black}\scriptsize }},
    scale=1.8
]

\pie{}
    {26.2/cwe416,19/cwe122,20.2/cwe787,19.6/cwe20,14.8/cwe78}

\legend[shift={(1.2cm,0.9cm)}]{{\small Use After Free}/cwe416}
\legend[shift={(1.2cm,0.6cm)}]{{\small Heap-based Buffer Overflow}/cwe122}
\legend[shift={(1.2cm,0.3cm)}]{{\small Out-of-bounds Write}/cwe787}
\legend[shift={(1.2cm,0.0cm)}]{ {\small Improper Input Validation}/cwe20}
\legend[shift={(1.2cm,-0.3cm)}]{{\small OS Command Injection}/cwe78}

\end{tikzpicture}
\vspace{-0.05in}
\caption{Top five firmware vulnerabilities (CVE)~\cite{cveorg}.}
\label{fig:cve-pi-chart}
\end{figure}

Figure~\ref{fig:cve-pi-chart} summarizes the proportional distribution of each category in the form of a pie chart. Our analysis reveals that approximately 38\% of all firmware-related CVEs are associated with memory corruption issues, specifically heap and stack-based buffer overflows. These memory-related flaws are particularly insidious due to their potential to enable arbitrary code execution, privilege escalation, and persistent backdoors, especially when present in low-level trusted code, such as bootloaders, device drivers, or management firmware.
Recent disclosures continue to confirm the prevalence and severity of these vulnerabilities. For example, CVE-2025-26336 (affecting Dell PowerEdge Chassis Management Controller) and CVE-2025-22457 (impacting Ivanti Connect Secure VPN Appliances) are both stack-based buffer overflow vulnerabilities that allow remote unauthenticated attackers to execute arbitrary code. Similarly, CVE-2025-23123, a heap buffer overflow vulnerability in Ubiquiti UniFi Protect Cameras, enables unauthorized network access and remote code execution through malformed requests. These examples illustrate that even in modern, high-assurance systems, such vulnerabilities persist and continue to pose a critical threat.

The continued emergence of these vulnerabilities highlights a broader issue: the inadequacy of traditional software testing and verification methods in addressing low-level firmware bugs. Tight coupling of firmware-hardware and its reliance on direct memory manipulation make it difficult to analyze using conventional static or dynamic analysis tools. This complexity enables even relatively simple memory bugs to escape detection during development and validation phases, ultimately propagating into deployed products.

\subsection{Overview of Coverage-driven Greybox Fuzzing}

Coverage-guided greybox fuzzing is one of the most effective techniques for automated vulnerability discovery. It operates by continuously generating and executing test inputs while using lightweight runtime feedback (such as code coverage) to guide exploration toward unexplored execution paths. Positioned between black-box and white-box testing, greybox fuzzing balances scalability with insight by gathering minimal yet informative execution data.

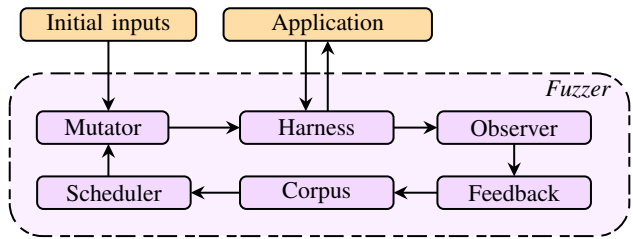
\begin{figure}[!h]
\centering
\small
\tikzset{every picture/.style={line width=0.75pt}} 

\begin{tikzpicture}[x=0.75pt,y=0.75pt,yscale=-1,xscale=1]

\draw  [fill={rgb, 255:red, 251; green, 240; blue, 255 }  ,fill opacity=1 ][dash pattern={on 3.75pt off 3pt on 7.5pt off 1.5pt}] (45,96.79) .. controls (45,87.71) and (52.36,80.36) .. (61.43,80.36) -- (341.07,80.36) .. controls (350.14,80.36) and (357.5,87.71) .. (357.5,96.79) -- (357.5,146.07) .. controls (357.5,155.14) and (350.14,162.5) .. (341.07,162.5) -- (61.43,162.5) .. controls (52.36,162.5) and (45,155.14) .. (45,146.07) -- cycle ;
\draw  [fill={rgb, 255:red, 244; green, 217; blue, 255 }  ,fill opacity=1 ] (260.71,102.81) .. controls (260.71,100.99) and (262.18,99.52) .. (263.99,99.52) -- (334.86,99.52) .. controls (336.67,99.52) and (338.14,100.99) .. (338.14,102.81) -- (338.14,112.67) .. controls (338.14,114.48) and (336.67,115.95) .. (334.86,115.95) -- (263.99,115.95) .. controls (262.18,115.95) and (260.71,114.48) .. (260.71,112.67) -- cycle ;
\draw  [fill={rgb, 255:red, 255; green, 221; blue, 166 }  ,fill opacity=1 ] (50.53,50.79) .. controls (50.53,48.97) and (52,47.5) .. (53.82,47.5) -- (135.74,47.5) .. controls (137.56,47.5) and (139.03,48.97) .. (139.03,50.79) -- (139.03,60.64) .. controls (139.03,62.46) and (137.56,63.93) .. (135.74,63.93) -- (53.82,63.93) .. controls (52,63.93) and (50.53,62.46) .. (50.53,60.64) -- cycle ;
\draw    (94.78,63.93) -- (94.78,96.52) ;
\draw [shift={(94.78,99.52)}, rotate = 270] [fill={rgb, 255:red, 0; green, 0; blue, 0 }  ][line width=0.08]  [draw opacity=0] (7.14,-3.43) -- (0,0) -- (7.14,3.43) -- (4.74,0) -- cycle    ;
\draw  [fill={rgb, 255:red, 244; green, 217; blue, 255 }  ,fill opacity=1 ] (58.67,102.81) .. controls (58.67,100.99) and (60.14,99.52) .. (61.96,99.52) -- (121.76,99.52) .. controls (123.57,99.52) and (125.04,100.99) .. (125.04,102.81) -- (125.04,112.67) .. controls (125.04,114.48) and (123.57,115.95) .. (121.76,115.95) -- (61.96,115.95) .. controls (60.14,115.95) and (58.67,114.48) .. (58.67,112.67) -- cycle ;
\draw  [fill={rgb, 255:red, 244; green, 217; blue, 255 }  ,fill opacity=1 ] (161.15,102.81) .. controls (161.15,100.99) and (162.62,99.52) .. (164.44,99.52) -- (235.3,99.52) .. controls (237.11,99.52) and (238.58,100.99) .. (238.58,102.81) -- (238.58,112.67) .. controls (238.58,114.48) and (237.11,115.95) .. (235.3,115.95) -- (164.44,115.95) .. controls (162.62,115.95) and (161.15,114.48) .. (161.15,112.67) -- cycle ;
\draw    (125.2,107.74) -- (158.15,107.74) ;
\draw [shift={(161.15,107.74)}, rotate = 180] [fill={rgb, 255:red, 0; green, 0; blue, 0 }  ][line width=0.08]  [draw opacity=0] (7.14,-3.43) -- (0,0) -- (7.14,3.43) -- (4.74,0) -- cycle    ;
\draw  [fill={rgb, 255:red, 255; green, 221; blue, 166 }  ,fill opacity=1 ] (152.85,50.79) .. controls (152.85,48.97) and (154.33,47.5) .. (156.14,47.5) -- (254.66,47.5) .. controls (256.47,47.5) and (257.94,48.97) .. (257.94,50.79) -- (257.94,60.64) .. controls (257.94,62.46) and (256.47,63.93) .. (254.66,63.93) -- (156.14,63.93) .. controls (154.33,63.93) and (152.85,62.46) .. (152.85,60.64) -- cycle ;
\draw    (194.34,63.93) -- (194.34,96.52) ;
\draw [shift={(194.34,99.52)}, rotate = 270] [fill={rgb, 255:red, 0; green, 0; blue, 0 }  ][line width=0.08]  [draw opacity=0] (7.14,-3.43) -- (0,0) -- (7.14,3.43) -- (4.74,0) -- cycle    ;
\draw  [fill={rgb, 255:red, 244; green, 217; blue, 255 }  ,fill opacity=1 ] (58.83,135.67) .. controls (58.83,133.85) and (60.3,132.38) .. (62.11,132.38) -- (132.98,132.38) .. controls (134.79,132.38) and (136.26,133.85) .. (136.26,135.67) -- (136.26,145.52) .. controls (136.26,147.34) and (134.79,148.81) .. (132.98,148.81) -- (62.11,148.81) .. controls (60.3,148.81) and (58.83,147.34) .. (58.83,145.52) -- cycle ;
\draw  [fill={rgb, 255:red, 244; green, 217; blue, 255 }  ,fill opacity=1 ] (260.71,135.67) .. controls (260.71,133.85) and (262.18,132.38) .. (263.99,132.38) -- (334.86,132.38) .. controls (336.67,132.38) and (338.14,133.85) .. (338.14,135.67) -- (338.14,145.52) .. controls (338.14,147.34) and (336.67,148.81) .. (334.86,148.81) -- (263.99,148.81) .. controls (262.18,148.81) and (260.71,147.34) .. (260.71,145.52) -- cycle ;
\draw  [fill={rgb, 255:red, 244; green, 217; blue, 255 }  ,fill opacity=1 ] (161.15,135.67) .. controls (161.15,133.85) and (162.62,132.38) .. (164.44,132.38) -- (235.3,132.38) .. controls (237.11,132.38) and (238.58,133.85) .. (238.58,135.67) -- (238.58,145.52) .. controls (238.58,147.34) and (237.11,148.81) .. (235.3,148.81) -- (164.44,148.81) .. controls (162.62,148.81) and (161.15,147.34) .. (161.15,145.52) -- cycle ;
\draw    (238.58,107.74) -- (257.71,107.74) ;
\draw [shift={(260.71,107.74)}, rotate = 180] [fill={rgb, 255:red, 0; green, 0; blue, 0 }  ][line width=0.08]  [draw opacity=0] (7.14,-3.43) -- (0,0) -- (7.14,3.43) -- (4.74,0) -- cycle    ;
\draw    (299.42,115.95) -- (299.42,129.38) ;
\draw [shift={(299.42,132.38)}, rotate = 270] [fill={rgb, 255:red, 0; green, 0; blue, 0 }  ][line width=0.08]  [draw opacity=0] (7.14,-3.43) -- (0,0) -- (7.14,3.43) -- (4.74,0) -- cycle    ;
\draw    (161.15,140.6) -- (139.26,140.6) ;
\draw [shift={(136.26,140.6)}, rotate = 360] [fill={rgb, 255:red, 0; green, 0; blue, 0 }  ][line width=0.08]  [draw opacity=0] (7.14,-3.43) -- (0,0) -- (7.14,3.43) -- (4.74,0) -- cycle    ;
\draw    (260.71,140.6) -- (241.58,140.6) ;
\draw [shift={(238.58,140.6)}, rotate = 360] [fill={rgb, 255:red, 0; green, 0; blue, 0 }  ][line width=0.08]  [draw opacity=0] (7.14,-3.43) -- (0,0) -- (7.14,3.43) -- (4.74,0) -- cycle    ;
\draw    (205.4,99.52) -- (205.4,66.93) ;
\draw [shift={(205.4,63.93)}, rotate = 90] [fill={rgb, 255:red, 0; green, 0; blue, 0 }  ][line width=0.08]  [draw opacity=0] (7.14,-3.43) -- (0,0) -- (7.14,3.43) -- (4.74,0) -- cycle    ;
\draw    (94.78,132.38) -- (94.78,118.95) ;
\draw [shift={(94.78,115.95)}, rotate = 90] [fill={rgb, 255:red, 0; green, 0; blue, 0 }  ][line width=0.08]  [draw opacity=0] (7.14,-3.43) -- (0,0) -- (7.14,3.43) -- (4.74,0) -- cycle    ;

\draw (299.42,107.74) node  [align=left] {\begin{minipage}[lt]{52.65pt}\setlength\topsep{0pt}
\begin{center}
Observer
\end{center}

\end{minipage}};
\draw (94.78,55.71) node [align=left] {\begin{minipage}[lt]{60.18pt}\setlength\topsep{0pt}
\begin{center}
Initial inputs
\end{center}

\end{minipage}};
\draw (92.01,107.74) node [align=left] {\begin{minipage}[lt]{45.13pt}\setlength\topsep{0pt}
\begin{center}
Mutator
\end{center}

\end{minipage}};
\draw (199.87,107.74) node  [align=left] {\begin{minipage}[lt]{52.65pt}\setlength\topsep{0pt}
\begin{center}
Harness
\end{center}

\end{minipage}};
\draw (205.4,55.71) node  [align=left] {\begin{minipage}[lt]{71.46pt}\setlength\topsep{0pt}
\begin{center}
Application
\end{center}

\end{minipage}};
\draw (97.54,140.6) node  [align=left] {\begin{minipage}[lt]{52.65pt}\setlength\topsep{0pt}
\begin{center}
Scheduler
\end{center}

\end{minipage}};
\draw (299.42,140.6) node [align=left] {\begin{minipage}[lt]{52.65pt}\setlength\topsep{0pt}
\begin{center}
Feedback
\end{center}

\end{minipage}};
\draw (199.87,140.6) node  [align=left] {\begin{minipage}[lt]{52.65pt}\setlength\topsep{0pt}
\begin{center}
Corpus
\end{center}

\end{minipage}};
\draw (309.47,82.21) node [anchor=north west][inner sep=0.75pt]   [align=left] {\textit{{ Fuzzer}}};

\end{tikzpicture}
\caption{An  overview of a greybox fuzzer}
\label{fig:fuzz_overview}
\end{figure}

Figure~\ref{fig:fuzz_overview} outlines the general workflow of a greybox fuzzer. The process begins with an initial seed corpus that is iteratively mutated through bit-level or structural transformations to produce new test cases. Each test case is executed through a harness that connects the fuzzer to the target application. The harness feeds the input and triggers execution. In firmware testing, it can delegate execution to an emulator or physical device, enabling interaction with low-level environments.

During execution, observers collect runtime metrics, such as branch coverage, crashes, or exceptions. These metrics are analyzed by a feedback engine to determine whether an input exercises novel behavior. Interesting inputs are added back to the corpus, while a scheduler prioritizes which cases to fuzz next based on criteria like coverage novelty or execution speed. This iterative, feedback-driven process continues until coverage converges, forming the foundation of most modern fuzzing frameworks.
While highly effective for conventional software, extending coverage-guided fuzzing to firmware remains challenging due to hardware dependencies, peripheral interactions, and privileged execution contexts.

\subsection{User-Level versus System-Level Emulation}\label{subsec:levels}

\begin{figure}[!h]
    \centering
    \small
    \tikzset{every picture/.style={line width=0.75pt}} 

\begin{tikzpicture}[x=0.75pt,y=0.75pt,yscale=-1,xscale=1]

\draw  [fill={rgb, 255:red, 255; green, 243; blue, 217 }  ,fill opacity=1 ] (24,107.2) .. controls (24,101.01) and (29.01,96) .. (35.2,96) -- (100.8,96) .. controls (106.99,96) and (112,101.01) .. (112,107.2) -- (112,152) .. controls (112,152) and (112,152) .. (112,152) -- (24,152) .. controls (24,152) and (24,152) .. (24,152) -- cycle ;
\draw  [fill={rgb, 255:red, 255; green, 178; blue, 105 }  ,fill opacity=1 ] (112.09,233.6) .. controls (112.1,237.13) and (109.25,240.01) .. (105.71,240.02) -- (30.57,240.28) .. controls (27.03,240.3) and (24.16,237.44) .. (24.15,233.91) -- (24.06,208.31) .. controls (24.06,208.31) and (24.06,208.31) .. (24.06,208.31) -- (112,208) .. controls (112,208) and (112,208) .. (112,208) -- cycle ;
\draw  [fill={rgb, 255:red, 255; green, 221; blue, 166 }  ,fill opacity=1 ] (24,152) -- (112,152) -- (112,208) -- (24,208) -- cycle ;
\draw  [fill={rgb, 255:red, 255; green, 243; blue, 217 }  ,fill opacity=1 ] (183.83,107.03) .. controls (183.83,100.85) and (188.85,95.83) .. (195.03,95.83) -- (260.63,95.83) .. controls (266.82,95.83) and (271.83,100.85) .. (271.83,107.03) -- (271.83,151.83) .. controls (271.83,151.83) and (271.83,151.83) .. (271.83,151.83) -- (183.83,151.83) .. controls (183.83,151.83) and (183.83,151.83) .. (183.83,151.83) -- cycle ;
\draw  [fill={rgb, 255:red, 255; green, 178; blue, 105 }  ,fill opacity=1 ] (271.98,233.29) .. controls (271.99,236.83) and (269.13,239.7) .. (265.6,239.72) -- (190.46,239.98) .. controls (186.92,239.99) and (184.05,237.13) .. (184.03,233.6) -- (183.94,208) .. controls (183.94,208) and (183.94,208) .. (183.94,208) -- (271.89,207.69) .. controls (271.89,207.69) and (271.89,207.69) .. (271.89,207.69) -- cycle ;
\draw  [fill={rgb, 255:red, 255; green, 221; blue, 166 }  ,fill opacity=1 ] (183.83,151.83) -- (271.83,151.83) -- (271.83,207.83) -- (183.83,207.83) -- cycle ;
\draw  [fill={rgb, 255:red, 255; green, 217; blue, 217 }  ,fill opacity=1 ][dash pattern={on 4.5pt off 4.5pt}] (192,134.23) .. controls (192,128.49) and (196.66,123.83) .. (202.4,123.83) -- (253.6,123.83) .. controls (259.34,123.83) and (264,128.49) .. (264,134.23) -- (264,165.43) .. controls (264,171.18) and (259.34,175.83) .. (253.6,175.83) -- (202.4,175.83) .. controls (196.66,175.83) and (192,171.18) .. (192,165.43) -- cycle ;
\draw  [fill={rgb, 255:red, 255; green, 217; blue, 217 }  ,fill opacity=1 ][dash pattern={on 4.5pt off 4.5pt}] (32,124.8) .. controls (32,122.15) and (34.15,120) .. (36.8,120) -- (99.2,120) .. controls (101.85,120) and (104,122.15) .. (104,124.8) -- (104,139.2) .. controls (104,141.85) and (101.85,144) .. (99.2,144) -- (36.8,144) .. controls (34.15,144) and (32,141.85) .. (32,139.2) -- cycle ;
\draw  [fill={rgb, 255:red, 255; green, 217; blue, 217 }  ,fill opacity=1 ][dash pattern={on 4.5pt off 4.5pt}] (32,164.8) .. controls (32,162.15) and (34.15,160) .. (36.8,160) -- (99.2,160) .. controls (101.85,160) and (104,162.15) .. (104,164.8) -- (104,179.2) .. controls (104,181.85) and (101.85,184) .. (99.2,184) -- (36.8,184) .. controls (34.15,184) and (32,181.85) .. (32,179.2) -- cycle ;
\draw  [dash pattern={on 4.5pt off 4.5pt}] (168,81.67) .. controls (168,76.33) and (172.33,72) .. (177.67,72) -- (278.33,72) .. controls (283.67,72) and (288,76.33) .. (288,81.67) -- (288,198.33) .. controls (288,203.67) and (283.67,208) .. (278.33,208) -- (177.67,208) .. controls (172.33,208) and (168,203.67) .. (168,198.33) -- cycle ;
\draw  [dash pattern={on 4.5pt off 4.5pt}] (8,78.44) .. controls (8,74.89) and (10.89,72) .. (14.44,72) -- (121.56,72) .. controls (125.11,72) and (128,74.89) .. (128,78.44) -- (128,145.56) .. controls (128,149.11) and (125.11,152) .. (121.56,152) -- (14.44,152) .. controls (10.89,152) and (8,149.11) .. (8,145.56) -- cycle ;
\draw  [draw opacity=0][fill={rgb, 255:red, 255; green, 255; blue, 255 }  ,fill opacity=1 ] (24,64) -- (112,64) -- (112,88) -- (24,88) -- cycle ;
\draw  [draw opacity=0][fill={rgb, 255:red, 255; green, 255; blue, 255 }  ,fill opacity=1 ] (184,63.5) -- (272,63.5) -- (272,87.5) -- (184,87.5) -- cycle ;

\draw (68,104) node  [font=\small] [align=left] {\begin{minipage}[lt]{59.84pt}\setlength\topsep{0pt}
\begin{center}
\textit{User level}
\end{center}

\end{minipage}};
\draw (68,200) node  [font=\small] [align=left] {\begin{minipage}[lt]{59.84pt}\setlength\topsep{0pt}
\begin{center}
\textit{Kernel level}
\end{center}

\end{minipage}};
\draw (68.17,232.01) node  [font=\small] [align=left] {\begin{minipage}[lt]{59.84pt}\setlength\topsep{0pt}
\begin{center}
\textit{Hardware}
\end{center}

\end{minipage}};
\draw (227.83,103.83) node  [font=\small] [align=left] {\begin{minipage}[lt]{59.84pt}\setlength\topsep{0pt}
\begin{center}
\textit{User level}
\end{center}

\end{minipage}};
\draw (228,149.83) node  [font=\footnotesize] [align=left] {\begin{minipage}[lt]{48.96pt}\setlength\topsep{0pt}
\begin{center}
\textbf{Firmware}
\end{center}

\end{minipage}};
\draw (68,132) node  [font=\footnotesize] [align=left] {\begin{minipage}[lt]{48.96pt}\setlength\topsep{0pt}
\begin{center}
\textbf{Program}
\end{center}

\end{minipage}};
\draw (68,172) node  [font=\footnotesize] [align=left] {\begin{minipage}[lt]{48.96pt}\setlength\topsep{0pt}
\begin{center}
\textbf{Kernel}
\end{center}

\end{minipage}};
\draw (228,231.83) node  [font=\small] [align=left] {\begin{minipage}[lt]{59.84pt}\setlength\topsep{0pt}
\begin{center}
\textit{Hardware}
\end{center}

\end{minipage}};
\draw (228,199.83) node  [font=\small] [align=left] {\begin{minipage}[lt]{59.84pt}\setlength\topsep{0pt}
\begin{center}
\textit{Kernel level}
\end{center}

\end{minipage}};
\draw (68,76) node   [align=left] {\begin{minipage}[lt]{60.18pt}\setlength\topsep{0pt}
\begin{center}
{\small User-level }\\{\small emulation}
\end{center}

\end{minipage}};
\draw (228,75.5) node   [align=left] {\begin{minipage}[lt]{60.18pt}\setlength\topsep{0pt}
\begin{center}
{\small System-level }\\{\small emulation}
\end{center}

\end{minipage}};

\end{tikzpicture}
    \caption{User-level versus system-level emulation}
    \label{fig:system_and_user_level}
\end{figure}
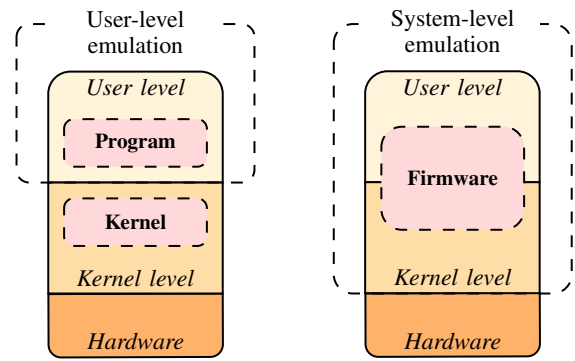

Firmware analysis relies heavily on emulation to execute binaries outside their native hardware environment. Figure~\ref{fig:system_and_user_level} shows two common approaches: user-level (or process-level) and system-level (or full-system) emulation.

In user-level emulation, the firmware binary is executed as an ordinary process on the host. The emulator dynamically translates target CPU instructions but relies on the host operating system to handle system calls and I/O. Only the user-space portion of the firmware runs in this mode; hardware peripherals, device drivers, and the kernel are excluded. This setup is lightweight and fast, making it suitable for rapid fuzzing or basic functional testing. However, it cannot accurately reproduce most real firmware environments. Firmware frequently interacts with kernel services, custom system libraries, and hardware components such as flash storage, network controllers, and memory-mapped I/O. As a result, user-level emulation often fails to execute firmware correctly or misses critical behavior that depends on privileged operations or hardware state~\cite{zheng2019firm, yin2021firmhunter}.

System-level emulation, in contrast, creates a complete virtual hardware platform for the firmware. The emulator models the CPU, memory, and peripheral devices, and boots the firmware’s actual kernel and user-space components inside a guest environment. System calls are handled by the guest kernel, and peripheral interactions are routed to virtual devices, allowing privileged firmware routines (e.g., device initialization, interrupt handling, and driver operations) to execute faithfully. This full-system context enables the analysis of low-level and hardware-dependent code paths that are invisible to user-level approaches. Overall, system-level emulation incurs higher computational overhead and resource usage, though techniques such as partial device modeling and selective instrumentation can mitigate this overhead up to some extent~\cite{zheng2019firm}.

\subsection{Related Work}
\label{sec:related_work}

Validating modern hardware platforms has become increasingly complex due to heterogeneous architectures, tightly coupled peripherals, and intricate hardware–software interactions that must behave correctly under diverse operating conditions~\cite{jayasena2024directed}. As this complexity shifts more functionality into firmware, ensuring the correctness and security of these low-level software layers has become equally critical. As we discussed earlier, firmware fuzzing has become a vital technique for discovering vulnerabilities in embedded systems. Over the years, fuzzing research has evolved from general-purpose software testing toward specialized frameworks that address the requirements demanded by firmware~\cite{zhang2020firmware, scharnowski2022fuzzware}.

Early greybox fuzzers, such as AFL~\cite{fioraldi2020afl++} and AFLFast~\cite{bohme2016coverage}, pioneered coverage-guided mutation strategies, enabling efficient exploration of user-space applications. Recent extensions, such as AFLNet~\cite{pham2020aflnet} and StateAFL~\cite{natella2022stateafl}, adapted these principles to stateful network protocols by incorporating protocol-aware feedback and memory snapshots. NSFuzz~\cite{qin2023nsfuzz} improved state modeling through variable-based representations. Although effective for conventional software, these techniques assume stable runtime environments and cannot directly handle hardware dependencies and low-level operations that are available in firmware.
To address this gap, several firmware-oriented fuzzers have been developed. FirmFuzz~\cite{zheng2019firm} introduced high-throughput greybox fuzzing via POSIX-compatible emulation, while FirmCOV~\cite{kim2021firm} enhanced coverage through process-level virtualization and structured-input dictionaries. FORMING~\cite{seidel2023forming} enabled near-native rehosting by executing firmware as Linux processes, and EMBER~\cite{farrelly2023ember} improved peripheral emulation through model-free handling of memory-mapped I/O. EM-Fuzz~\cite{gao2020fuzz} further augmented firmware fuzzing with memory-access monitoring to detect memory-sensitive vulnerabilities. Despite these advances, most of these methods operate at the user or process level, leaving kernel-space and hardware-interaction vulnerabilities largely unexplored.

Efforts to increase fuzzing efficiency have led to snapshot-based methods, such as SNPSFuzzer~\cite{li2022snpsfuzzer}, which saves program states to avoid repetitive initialization, and Nyx~\cite{schumilo2022nyx}, which employs hypervisor-level snapshots to accelerate fuzzing. Similarly, SnapFuzz~\cite{andronidis2022snapfuzz} synchronizes asynchronous I/O through in-memory file systems. While these approaches improve throughput, they do not overcome the fundamental coverage stagnation caused by complex, constrained branches common in firmware execution paths.

Hybrid fuzzing frameworks integrate the scalability of fuzzing with the path-sensitive precision of symbolic execution. Early efforts, such as Driller~\cite{stephens2016driller}, demonstrated that selectively invoking symbolic execution when fuzzing progress stagnates can improve coverage in complex programs. Later, QSYM~\cite{yun2018qsym} introduced a more efficient symbolic execution engine tailored for x86-64 binaries. However, these frameworks were primarily developed for user-space applications and lack the capability to emulate or analyze full firmware environments. In parallel, hardware fuzzing approaches have adopted similar hybrid strategies, leveraging symbolic execution to enhance coverage in processor and RTL testing~\cite{jayasena2025fuss}. Recent approaches, including LLM4Fuzz~\cite{meng2024large} and LLMIF~\cite{wang2024llmif}, explore the use of large language models (LLMs) to improve input diversity through grammar inference, though these techniques are unsuitable for low-level firmware analysis, where hardware interaction and state modeling are essential. Complementing these efforts, FirmWall~\cite{jayasena2025firmwall} employs directed symbolic execution to analyze system-call–relevant paths in firmware binaries, enabling the detection of unauthorized or policy-violating interactions with underlying hardware; however, its analysis remains constrained by the complexity of the firmware due to state space explosion.

In summary, existing research predominantly focuses on user-level or partially emulated firmware, abstracting away hardware interactions that are critical to comprehensive vulnerability detection. Such abstractions risk overlooking flaws that arise only during kernel-space execution or direct hardware communication. This highlights the need for system-level fuzzers that can handle both user-level and system-level firmware. Our proposed framework, SysFuSS, provides a system-level fuzzing framework that can efficiently detect both user-level and system-level firmware vulnerabilities.

\section{SysFuSS Methodology}\label{sec:methodology}

\begin{figure}[htb]
    \centering
    \small
    \tikzset{every picture/.style={line width=0.75pt}} 

\begin{tikzpicture}[x=0.75pt,y=0.75pt,yscale=-1,xscale=1]

\draw  [fill={rgb, 255:red, 255; green, 243; blue, 217 }  ,fill opacity=1 ] (255,170) -- (355,170) -- (355,295) -- (255,295) -- cycle ;
\draw  [fill={rgb, 255:red, 255; green, 243; blue, 217 }  ,fill opacity=1 ] (75,170) -- (175,170) -- (175,295) -- (75,295) -- cycle ;
\draw  [fill={rgb, 255:red, 255; green, 221; blue, 166 }  ,fill opacity=1 ] (75,118) .. controls (75,113.58) and (78.58,110) .. (83,110) -- (347,110) .. controls (351.42,110) and (355,113.58) .. (355,118) -- (355,142) .. controls (355,146.42) and (351.42,150) .. (347,150) -- (83,150) .. controls (78.58,150) and (75,146.42) .. (75,142) -- cycle ;
\draw  [fill={rgb, 255:red, 255; green, 178; blue, 105 }  ,fill opacity=1 ] (80,226) .. controls (80,222.69) and (82.69,220) .. (86,220) -- (164,220) .. controls (167.31,220) and (170,222.69) .. (170,226) -- (170,244) .. controls (170,247.31) and (167.31,250) .. (164,250) -- (86,250) .. controls (82.69,250) and (80,247.31) .. (80,244) -- cycle ;
\draw  [fill={rgb, 255:red, 255; green, 178; blue, 105 }  ,fill opacity=1 ] (80,266) .. controls (80,262.69) and (82.69,260) .. (86,260) -- (164,260) .. controls (167.31,260) and (170,262.69) .. (170,266) -- (170,284) .. controls (170,287.31) and (167.31,290) .. (164,290) -- (86,290) .. controls (82.69,290) and (80,287.31) .. (80,284) -- cycle ;
\draw  [fill={rgb, 255:red, 255; green, 221; blue, 166 }  ,fill opacity=1 ] (75,326) .. controls (75,319.92) and (79.92,315) .. (86,315) -- (344,315) .. controls (350.08,315) and (355,319.92) .. (355,326) -- (355,359) .. controls (355,365.08) and (350.08,370) .. (344,370) -- (86,370) .. controls (79.92,370) and (75,365.08) .. (75,359) -- cycle ;
\draw    (215,84.61) -- (215,106.61) ;
\draw [shift={(215,109.61)}, rotate = 270] [fill={rgb, 255:red, 0; green, 0; blue, 0 }  ][line width=0.08]  [draw opacity=0] (8.93,-4.29) -- (0,0) -- (8.93,4.29) -- (5.93,0) -- cycle    ;
\draw    (305,150) -- (305,167) ;
\draw [shift={(305,170)}, rotate = 270] [fill={rgb, 255:red, 0; green, 0; blue, 0 }  ][line width=0.08]  [draw opacity=0] (8.93,-4.29) -- (0,0) -- (8.93,4.29) -- (5.93,0) -- cycle    ;
\draw    (305,315) -- (305,298) ;
\draw [shift={(305,295)}, rotate = 90] [fill={rgb, 255:red, 0; green, 0; blue, 0 }  ][line width=0.08]  [draw opacity=0] (8.93,-4.29) -- (0,0) -- (8.93,4.29) -- (5.93,0) -- cycle    ;
\draw    (125,150) -- (125,167) ;
\draw [shift={(125,170)}, rotate = 270] [fill={rgb, 255:red, 0; green, 0; blue, 0 }  ][line width=0.08]  [draw opacity=0] (8.93,-4.29) -- (0,0) -- (8.93,4.29) -- (5.93,0) -- cycle    ;
\draw  [fill={rgb, 255:red, 255; green, 178; blue, 105 }  ,fill opacity=1 ] (260,226) .. controls (260,222.69) and (262.69,220) .. (266,220) -- (344,220) .. controls (347.31,220) and (350,222.69) .. (350,226) -- (350,244) .. controls (350,247.31) and (347.31,250) .. (344,250) -- (266,250) .. controls (262.69,250) and (260,247.31) .. (260,244) -- cycle ;
\draw  [fill={rgb, 255:red, 255; green, 221; blue, 166 }  ,fill opacity=1 ] (235,195.63) -- (235,229.38) .. controls (235,232.48) and (226.05,235) .. (215,235) .. controls (203.95,235) and (195,232.48) .. (195,229.38) -- (195,195.63) .. controls (195,192.52) and (203.95,190) .. (215,190) .. controls (226.05,190) and (235,192.52) .. (235,195.63) .. controls (235,198.73) and (226.05,201.25) .. (215,201.25) .. controls (203.95,201.25) and (195,198.73) .. (195,195.63) ;
\draw  [fill={rgb, 255:red, 222; green, 222; blue, 255 }  ,fill opacity=1 ] (187,45) -- (255,45) -- (255,74.7) .. controls (212.5,74.7) and (221,85.41) .. (187,78.48) -- cycle ; \draw  [fill={rgb, 255:red, 222; green, 222; blue, 255 }  ,fill opacity=1 ] (178.5,49.5) -- (246.5,49.5) -- (246.5,79.2) .. controls (204,79.2) and (212.5,89.91) .. (178.5,82.98) -- cycle ; \draw  [fill={rgb, 255:red, 222; green, 222; blue, 255 }  ,fill opacity=1 ] (170,54) -- (238,54) -- (238,83.7) .. controls (195.5,83.7) and (204,94.41) .. (170,87.48) -- cycle ;

\draw    (255,225) -- (238,225) ;
\draw [shift={(235,225)}, rotate = 360] [fill={rgb, 255:red, 0; green, 0; blue, 0 }  ][line width=0.08]  [draw opacity=0] (8.93,-4.29) -- (0,0) -- (8.93,4.29) -- (5.93,0) -- cycle    ;
\draw    (195,225) -- (178,225) ;
\draw [shift={(175,225)}, rotate = 360] [fill={rgb, 255:red, 0; green, 0; blue, 0 }  ][line width=0.08]  [draw opacity=0] (8.93,-4.29) -- (0,0) -- (8.93,4.29) -- (5.93,0) -- cycle    ;
\draw    (175,205) -- (192,205) ;
\draw [shift={(195,205)}, rotate = 180] [fill={rgb, 255:red, 0; green, 0; blue, 0 }  ][line width=0.08]  [draw opacity=0] (8.93,-4.29) -- (0,0) -- (8.93,4.29) -- (5.93,0) -- cycle    ;
\draw    (235,205) -- (252,205) ;
\draw [shift={(255,205)}, rotate = 180] [fill={rgb, 255:red, 0; green, 0; blue, 0 }  ][line width=0.08]  [draw opacity=0] (8.93,-4.29) -- (0,0) -- (8.93,4.29) -- (5.93,0) -- cycle    ;
\draw    (125,295) -- (125,312) ;
\draw [shift={(125,315)}, rotate = 270] [fill={rgb, 255:red, 0; green, 0; blue, 0 }  ][line width=0.08]  [draw opacity=0] (8.93,-4.29) -- (0,0) -- (8.93,4.29) -- (5.93,0) -- cycle    ;
\draw  [fill={rgb, 255:red, 255; green, 178; blue, 105 }  ,fill opacity=1 ] (230,341) .. controls (230,337.69) and (232.69,335) .. (236,335) -- (334,335) .. controls (337.31,335) and (340,337.69) .. (340,341) -- (340,359) .. controls (340,362.31) and (337.31,365) .. (334,365) -- (236,365) .. controls (232.69,365) and (230,362.31) .. (230,359) -- cycle ;
\draw  [fill={rgb, 255:red, 255; green, 178; blue, 105 }  ,fill opacity=1 ] (90,341) .. controls (90,337.69) and (92.69,335) .. (96,335) -- (199,335) .. controls (202.31,335) and (205,337.69) .. (205,341) -- (205,359) .. controls (205,362.31) and (202.31,365) .. (199,365) -- (96,365) .. controls (92.69,365) and (90,362.31) .. (90,359) -- cycle ;
\draw  [fill={rgb, 255:red, 255; green, 178; blue, 105 }  ,fill opacity=1 ] (260,266) .. controls (260,262.69) and (262.69,260) .. (266,260) -- (344,260) .. controls (347.31,260) and (350,262.69) .. (350,266) -- (350,284) .. controls (350,287.31) and (347.31,290) .. (344,290) -- (266,290) .. controls (262.69,290) and (260,287.31) .. (260,284) -- cycle ;

\draw (215,130) node   [align=left] {\begin{minipage}[lt]{190.4pt}\setlength\topsep{0pt}
\begin{center}
{\footnotesize \textbf{Design Instrumentaion}}\\{\footnotesize \textit{(Section 3.1)}}
\end{center}

\end{minipage}};
\draw (125.5,192) node   [align=left] {\begin{minipage}[lt]{68pt}\setlength\topsep{0pt}
\begin{center}
{\footnotesize \textbf{Fuzzing and Plateau Detection}}\\{\footnotesize \textit{(Section 3.2)}}
\end{center}

\end{minipage}};
\draw (125,235) node   [align=left] {\begin{minipage}[lt]{61.2pt}\setlength\topsep{0pt}
\begin{center}
{\footnotesize \textit{Coverage Analyzer}}
\end{center}

\end{minipage}};
\draw (124.75,275.13) node   [align=left] {\begin{minipage}[lt]{60.86pt}\setlength\topsep{0pt}
\begin{center}
{\footnotesize \textit{Plateau Detection}}
\end{center}

\end{minipage}};
\draw (215,325) node   [align=left] {\begin{minipage}[lt]{190.4pt}\setlength\topsep{0pt}
\begin{center}
{\footnotesize \textbf{Constraint Generator }\textit{(Section 3.3)}}
\end{center}

\end{minipage}};
\draw (305,192.5) node   [align=left] {\begin{minipage}[lt]{68pt}\setlength\topsep{0pt}
\begin{center}
{\footnotesize \textbf{Selective Symbolic Execution}}\\{\footnotesize \textit{(Section 3.4)}}
\end{center}

\end{minipage}};
\draw (305,235) node   [align=left] {\begin{minipage}[lt]{61.2pt}\setlength\topsep{0pt}
\begin{center}
{\footnotesize \textit{Symbolic Initialization}}
\end{center}

\end{minipage}};
\draw (204,68.85) node   [align=left] {\begin{minipage}[lt]{46.24pt}\setlength\topsep{0pt}
\begin{center}
{\footnotesize Firmware}
\end{center}

\end{minipage}};
\draw (215,212.5) node   [align=left] {\begin{minipage}[lt]{27.2pt}\setlength\topsep{0pt}
\begin{center}
{\footnotesize \textbf{Corpus}}
\end{center}

\end{minipage}};
\draw (285,350) node   [align=left] {\begin{minipage}[lt]{74.8pt}\setlength\topsep{0pt}
\begin{center}
{\footnotesize \textit{Prioritization}}
\end{center}

\end{minipage}};
\draw (147.5,350) node   [align=left] {\begin{minipage}[lt]{78.2pt}\setlength\topsep{0pt}
\begin{center}
{\footnotesize \textit{Unexplored Path Identification}}
\end{center}

\end{minipage}};
\draw (305,275) node   [align=left] {\begin{minipage}[lt]{61.2pt}\setlength\topsep{0pt}
\begin{center}
{\footnotesize \textit{Path Solver}}
\end{center}

\end{minipage}};

\end{tikzpicture}
    \caption{Overview of SysFuSS that includes five components: design instrumentation, fuzzing, coverage plateau detection, constraint generation, and symbolic execution.}
    \label{fig:overview}
\end{figure}
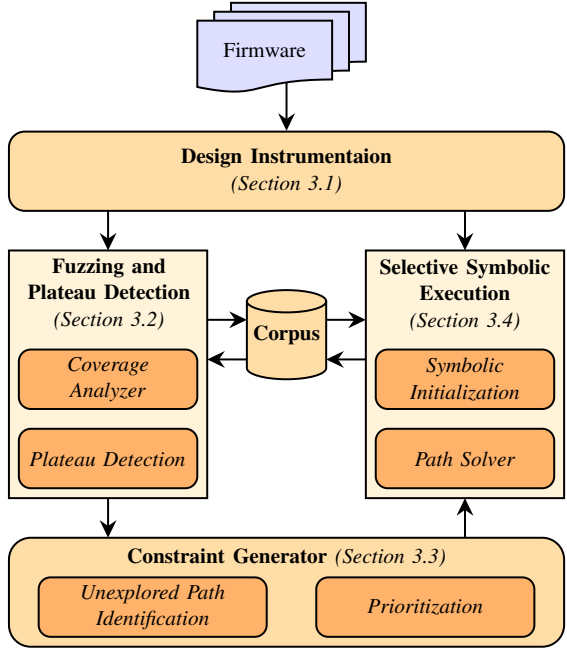

Figure~\ref{fig:overview} shows an overview of our proposed (SysFuSS) framework that consists of five major components: (i) firmware instrumentation for coverage tracking, (ii) fuzzing execution on both user level and system level, (iii) automatic coverage plateau detection, (iv) constraint extraction from hard-to-reach code regions, and (v) selective symbolic execution for generation of directed tests.  
Algorithm~\ref{alg:sysfuss} shows the five major components outlined in Figure~\ref{fig:overview} and how they interact with each other. The following subsections describe each of these components in detail.

\begin{algorithm}[t]
\caption{Fuzzing with Selective Symbolic Execution}
\label{alg:sysfuss}
\begin{algorithmic}[1]
\INPUT Firmware design $\mathcal{F}$, Initial corpus $\mathcal{C}$, Plateau window $W$, Plateau threshold $\epsilon$, Max constraint pairs $L$, Test time $T_{time}$
\OUTPUT Test input set $\mathcal{T}$

\STATE $\mathcal{T} \leftarrow \mathcal{C}$ \COMMENT{Initialize test set with corpus}
\STATE $C(0) \leftarrow 0$ \COMMENT{Initial coverage is zero}
\STATE $i \leftarrow 1$ \COMMENT{Initialize iteration counter}
\STATE $\mathcal{G} \leftarrow \textsc{InstrumentDesign}(\mathcal{F})$ \COMMENT{Section 3.1}

\WHILE{$time<T_{time}$}
    \STATE $\mathcal{T}_{new} \leftarrow \textsc{RunFuzzer}(\mathcal{G}, \mathcal{T})$ \COMMENT{Section 3.2}
    \STATE $C(t) \leftarrow \textsc{MeasureCoverage}(\mathcal{G}, \mathcal{T} \cup \mathcal{T}_{new})$ 
    \IF{$i \geq W$ \AND $\frac{C(i) - C(i-W)}{W} < \epsilon$}
        \STATE $\mathit{CP} \leftarrow \textsc{GetConstraints}(\mathit{CFG})$ \COMMENT{Section 3.3}

        \STATE $Q \leftarrow \textsc{MinHeap}(\mathit{CP})$
        \STATE $Q \leftarrow \textsc{Prune}(Q, L)$ \COMMENT{Limit queue size to $L$}
        
        \FORALL{$(src, dst) \in Q$}
            \STATE $\Sigma \leftarrow \textsc{StatePrep}(\mathcal{G}, \mathcal{T}_{new}, src)$ 
            \STATE $\mathit{I}_{sym} \leftarrow \textsc{MarkSymbolic}(\Sigma, src)$
            \STATE $\phi_{src \to dst} \leftarrow \textsc{RunSymbolic}(\Sigma, src, dst, \mathit{I}_{sym})$ 
            
            \IF{$\textsc{SolveSMT}(\phi_{src \to dst}) \neq \emptyset$}
                \STATE $I_{c} \leftarrow \textsc{SolveSMT}(\phi_{src \to dst})$ 
                \STATE $\mathcal{T} \leftarrow \mathcal{T} \cup \{I_{c}\}$ \COMMENT{Add to corpus}
            \ENDIF
        \ENDFOR
    \ELSE
        \STATE $\mathcal{T} \leftarrow \mathcal{T} \cup \mathcal{T}_{new}$ \COMMENT{Continue fuzzing, add new tests}
    \ENDIF
    
    \STATE $i \leftarrow i + 1$ \COMMENT{Increment iteration counter}
\ENDWHILE

\RETURN $\mathcal{T}$
\end{algorithmic}
\end{algorithm}

\subsection{Design Instrumentation}

The primary goal of SysFuSS is to uncover firmware vulnerabilities while maximizing code coverage. Achieving this requires instrumenting the firmware so that both the fuzzing and symbolic execution engines can interpret the program structure and share execution context. To this end, we first construct a control-flow graph (CFG) of the firmware binary, then embed coverage trackers and insert code hooks to interface with the fuzzing engine. Finally, we integrate memory instrumentation to detect potential memory corruption vulnerabilities during execution. This four-step process corresponds to the line 4 ({\sc InstrumentDesign}) in Algorithm~\ref{alg:sysfuss}. The remainder of this section describes these four steps in detail.

\subsubsection{Control Flow Graph (CFG) Construction}
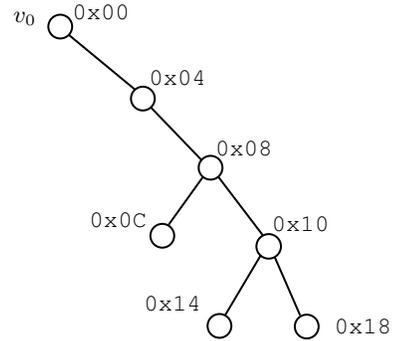
\begin{figure}[!htb]
    \centering
    \small
    \tikzset{every picture/.style={line width=0.75pt}} 

\begin{tikzpicture}[x=0.75pt,y=0.75pt,yscale=-1,xscale=1]

\draw  [color={rgb, 255:red, 0; green, 0; blue, 0 }  ,draw opacity=1 ][line width=0.75]  (272.51,104.33) .. controls (272.51,101.02) and (275.2,98.33) .. (278.51,98.33) .. controls (281.82,98.33) and (284.51,101.02) .. (284.51,104.33) .. controls (284.51,107.65) and (281.82,110.33) .. (278.51,110.33) .. controls (275.2,110.33) and (272.51,107.65) .. (272.51,104.33) -- cycle ;
\draw  [color={rgb, 255:red, 0; green, 0; blue, 0 }  ,draw opacity=1 ][line width=0.75]  (306.64,139.23) .. controls (306.64,135.92) and (309.33,133.23) .. (312.64,133.23) .. controls (315.96,133.23) and (318.64,135.92) .. (318.64,139.23) .. controls (318.64,142.55) and (315.96,145.23) .. (312.64,145.23) .. controls (309.33,145.23) and (306.64,142.55) .. (306.64,139.23) -- cycle ;
\draw  [color={rgb, 255:red, 0; green, 0; blue, 0 }  ,draw opacity=1 ][line width=0.75]  (335.77,178.92) .. controls (335.77,175.5) and (338.54,172.73) .. (341.96,172.73) .. controls (345.37,172.73) and (348.14,175.5) .. (348.14,178.92) .. controls (348.14,182.34) and (345.37,185.11) .. (341.96,185.11) .. controls (338.54,185.11) and (335.77,182.34) .. (335.77,178.92) -- cycle ;
\draw    (282.5,108.75) -- (308,135.25) ;
\draw    (308.81,143.9) -- (291.51,168) ;
\draw    (316.5,143.75) -- (339.01,173.33) ;
\draw    (338.01,183.33) -- (320.01,213.67) ;
\draw    (357.81,213.73) -- (345.01,184.33) ;
\draw  [color={rgb, 255:red, 0; green, 0; blue, 0 }  ,draw opacity=1 ][line width=0.75]  (282.31,173.57) .. controls (282.31,170.25) and (285,167.57) .. (288.31,167.57) .. controls (291.62,167.57) and (294.31,170.25) .. (294.31,173.57) .. controls (294.31,176.88) and (291.62,179.57) .. (288.31,179.57) .. controls (285,179.57) and (282.31,176.88) .. (282.31,173.57) -- cycle ;
\draw  [color={rgb, 255:red, 0; green, 0; blue, 0 }  ,draw opacity=1 ][line width=0.75]  (311.14,219.07) .. controls (311.14,215.75) and (313.83,213.07) .. (317.14,213.07) .. controls (320.46,213.07) and (323.14,215.75) .. (323.14,219.07) .. controls (323.14,222.38) and (320.46,225.07) .. (317.14,225.07) .. controls (313.83,225.07) and (311.14,222.38) .. (311.14,219.07) -- cycle ;
\draw  [color={rgb, 255:red, 0; green, 0; blue, 0 }  ,draw opacity=1 ][line width=0.75]  (355.14,219.4) .. controls (355.14,216.09) and (357.83,213.4) .. (361.14,213.4) .. controls (364.46,213.4) and (367.14,216.09) .. (367.14,219.4) .. controls (367.14,222.71) and (364.46,225.4) .. (361.14,225.4) .. controls (357.83,225.4) and (355.14,222.71) .. (355.14,219.4) -- cycle ;
\draw  [color={rgb, 255:red, 0; green, 0; blue, 0 }  ,draw opacity=1 ][line width=0.75]  (230.84,68) .. controls (230.84,64.69) and (233.53,62) .. (236.84,62) .. controls (240.16,62) and (242.84,64.69) .. (242.84,68) .. controls (242.84,71.31) and (240.16,74) .. (236.84,74) .. controls (233.53,74) and (230.84,71.31) .. (230.84,68) -- cycle ;
\draw    (241.34,72.33) -- (274.5,99.75) ;

\draw (258.18,60.75) node   [align=left] {\begin{minipage}[lt]{22.89pt}\setlength\topsep{0pt}
{\fontfamily{pcr}\selectfont 0x00}
\end{minipage}};
\draw (296.84,94) node   [align=left] {\begin{minipage}[lt]{22.89pt}\setlength\topsep{0pt}
{\fontfamily{pcr}\selectfont 0x04}
\end{minipage}};
\draw (330.31,129.15) node   [align=left] {\begin{minipage}[lt]{22.89pt}\setlength\topsep{0pt}
{\fontfamily{pcr}\selectfont 0x08}
\end{minipage}};
\draw (266.81,165.82) node   [align=left] {\begin{minipage}[lt]{22.89pt}\setlength\topsep{0pt}
{\fontfamily{pcr}\selectfont 0x0C}
\end{minipage}};
\draw (358.79,167.34) node   [align=left] {\begin{minipage}[lt]{22.89pt}\setlength\topsep{0pt}
{\fontfamily{pcr}\selectfont 0x10}
\end{minipage}};
\draw (294.31,207.48) node   [align=left] {\begin{minipage}[lt]{22.89pt}\setlength\topsep{0pt}
{\fontfamily{pcr}\selectfont 0x14}
\end{minipage}};
\draw (390.34,219) node   [align=left] {\begin{minipage}[lt]{22.89pt}\setlength\topsep{0pt}
{\fontfamily{pcr}\selectfont 0x18}
\end{minipage}};
\draw (226.34,63.88) node   [align=left] {\begin{minipage}[lt]{19.72pt}\setlength\topsep{0pt}
$\displaystyle v_{0}$
\end{minipage}};

\end{tikzpicture}
    \caption{Control Flow Graph (CFG) constructed for the firmware in Figure~\ref{lst:ex_1}. Basic blocks are represented in white circles. The \textit{entry point} to the CFG is marked as $v_0$}
    \label{fig:cfg}
    \vspace{-0.5cm}
\end{figure}

We begin by analyzing the compiled firmware binary to construct a comprehensive CFG that represents the program's structure. The CFG serves as a fundamental data structure for tracking code coverage and guiding both fuzzing and symbolic execution. Each node in the CFG corresponds to a basic block, a sequence of instructions with a single entry point and a single exit point. Edges represent control flow transfers between basic blocks, including both direct jumps and conditional branches. An example CFG is shown in ~\ref{fig:cfg}.
To build the CFG, we employ static binary analysis techniques on the firmware image. This CFG is shared between the fuzzing and symbolic execution engines, enabling seamless handoff of coverage information when switching between the two techniques. During fuzzing, the CFG is updated dynamically to reflect which basic blocks and edges have been explored. When it reaches a coverage plateau, the symbolic execution engine consults the CFG to identify unexplored branches and generate inputs targeting those specific paths. 

\begin{figure}[t]
    \centering
    \fbox{\includegraphics[width=0.97\linewidth]{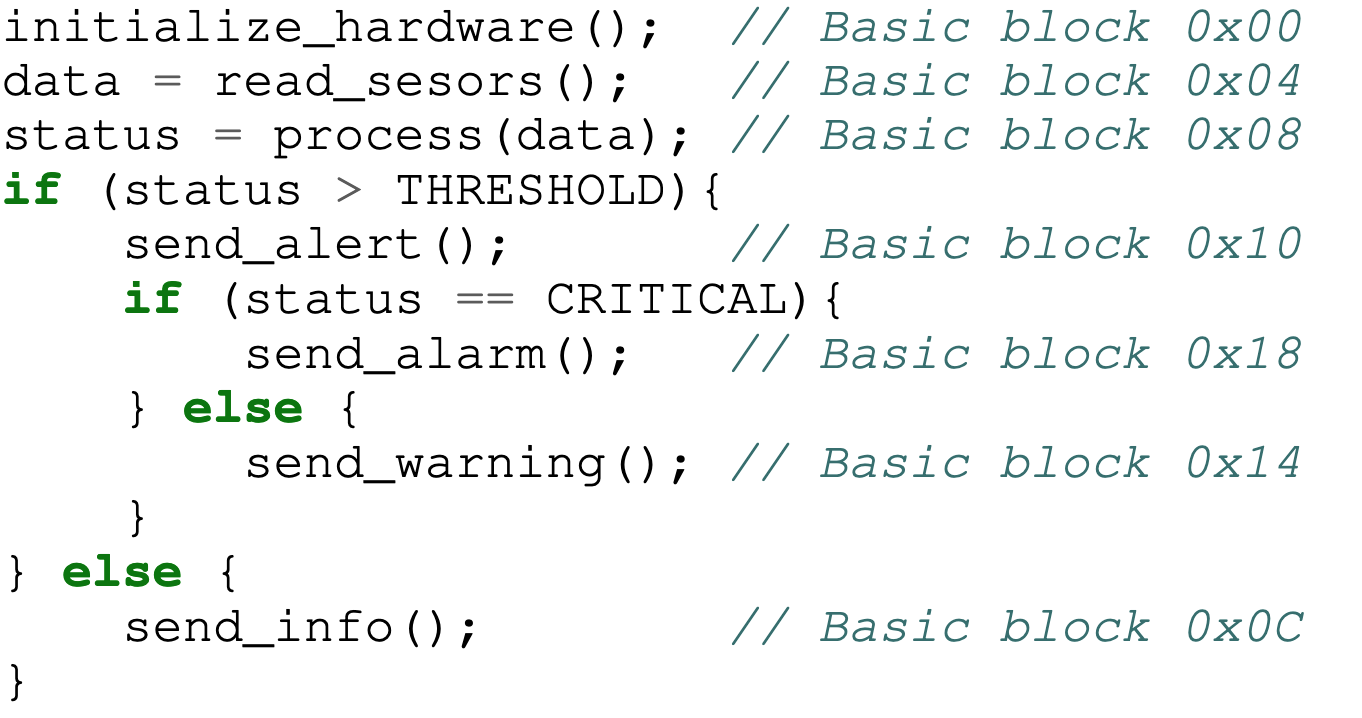}}
    \caption{A sample firmware code snippet}
    \label{lst:ex_1}
\end{figure}

{
\vspace{0.05in}
\noindent \underline{\textbf{Example 1} (CFG Construction)}: Figure~\ref{fig:cfg} illustrated the CFG constructed for the sample firmware shown in Figure~\ref{lst:ex_1}. In this CFG, each node represents a basic block labeled with its starting memory address. Execution begins at 0x00 and flows sequentially through 0x04 to 0x08. At 0x08, a conditional statement creates a branch, splitting execution into two paths: one to 0x0C and another to 0x10. Block 0x10 contains another conditional, further branching to 0x14 and 0x18. The edges between nodes represent possible execution paths through the program.
\exampleEnd
}



\subsubsection{Coverage Instrumentation}
To enable coverage-guided fuzzing, we instrument the firmware binary to track execution at the basic block level. This instrumentation records which blocks are executed during each test case, providing the feedback mechanism essential for fuzzing. When a basic block executes, the corresponding node in the CFG is marked as covered, allowing the fuzzer to quickly determine whether a test case has discovered new area in the CFG. We adopt the edge coverage approach of AFL~\cite{fioraldi2020afl++}, where we track transitions between basic blocks rather than block execution. This provides finer-grained feedback and helps the fuzzer discover complex execution paths effectively.

{
\vspace{0.05in}
\noindent \underline{\textbf{Example 2} (Coverage Instrumentation)}: To track code coverage during fuzzing, a colorization technique is applied to the CFG. In this example, the first fuzzing iteration executes the true path of both conditional statements, traversing blocks 0x00 → 0x04 → 0x08 → 0x10 → 0x18. These executed blocks are colored (shown in orange) in the CFG, while unvisited blocks (0x0C and 0x14) remain uncolored. This provides code coverage results, allowing us to identify which basic blocks have been executed and which remain unexplored during the fuzzing process.
\exampleEnd
}

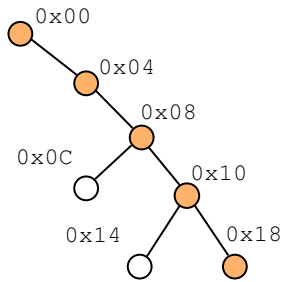
\begin{figure}[!htb]
    \centering
    \small
    \tikzset{every picture/.style={line width=0.75pt}} 

\begin{tikzpicture}[x=0.75pt,y=0.75pt,yscale=-1,xscale=1]

\draw  [color={rgb, 255:red, 0; green, 0; blue, 0 }  ,draw opacity=1 ][fill={rgb, 255:red, 255; green, 178; blue, 105 }  ,fill opacity=1 ][line width=0.75]  (280.5,99.5) .. controls (280.5,96.19) and (283.19,93.5) .. (286.5,93.5) .. controls (289.81,93.5) and (292.5,96.19) .. (292.5,99.5) .. controls (292.5,102.81) and (289.81,105.5) .. (286.5,105.5) .. controls (283.19,105.5) and (280.5,102.81) .. (280.5,99.5) -- cycle ;
\draw  [color={rgb, 255:red, 0; green, 0; blue, 0 }  ,draw opacity=1 ][fill={rgb, 255:red, 255; green, 178; blue, 105 }  ,fill opacity=1 ][line width=0.75]  (308.5,126.83) .. controls (308.5,123.52) and (311.19,120.83) .. (314.5,120.83) .. controls (317.81,120.83) and (320.5,123.52) .. (320.5,126.83) .. controls (320.5,130.15) and (317.81,132.83) .. (314.5,132.83) .. controls (311.19,132.83) and (308.5,130.15) .. (308.5,126.83) -- cycle ;
\draw  [color={rgb, 255:red, 0; green, 0; blue, 0 }  ,draw opacity=1 ][fill={rgb, 255:red, 255; green, 178; blue, 105 }  ,fill opacity=1 ][line width=0.75]  (331.13,156.19) .. controls (331.13,152.77) and (333.9,150) .. (337.31,150) .. controls (340.73,150) and (343.5,152.77) .. (343.5,156.19) .. controls (343.5,159.6) and (340.73,162.38) .. (337.31,162.38) .. controls (333.9,162.38) and (331.13,159.6) .. (331.13,156.19) -- cycle ;
\draw    (291,102.83) -- (310,122.17) ;
\draw    (309.33,130.83) -- (291,147.83) ;
\draw    (318,131.75) -- (334,151.25) ;
\draw    (333.83,161) -- (316.83,187.33) ;
\draw    (358.17,186.33) -- (341.17,161) ;
\draw  [color={rgb, 255:red, 0; green, 0; blue, 0 }  ,draw opacity=1 ][line width=0.75]  (280.5,152.5) .. controls (280.5,149.19) and (283.19,146.5) .. (286.5,146.5) .. controls (289.81,146.5) and (292.5,149.19) .. (292.5,152.5) .. controls (292.5,155.81) and (289.81,158.5) .. (286.5,158.5) .. controls (283.19,158.5) and (280.5,155.81) .. (280.5,152.5) -- cycle ;
\draw  [color={rgb, 255:red, 0; green, 0; blue, 0 }  ,draw opacity=1 ][line width=0.75]  (307.5,192) .. controls (307.5,188.69) and (310.19,186) .. (313.5,186) .. controls (316.81,186) and (319.5,188.69) .. (319.5,192) .. controls (319.5,195.31) and (316.81,198) .. (313.5,198) .. controls (310.19,198) and (307.5,195.31) .. (307.5,192) -- cycle ;
\draw  [color={rgb, 255:red, 0; green, 0; blue, 0 }  ,draw opacity=1 ][fill={rgb, 255:red, 255; green, 178; blue, 105 }  ,fill opacity=1 ][line width=0.75]  (355.5,192) .. controls (355.5,188.69) and (358.19,186) .. (361.5,186) .. controls (364.81,186) and (367.5,188.69) .. (367.5,192) .. controls (367.5,195.31) and (364.81,198) .. (361.5,198) .. controls (358.19,198) and (355.5,195.31) .. (355.5,192) -- cycle ;
\draw  [color={rgb, 255:red, 0; green, 0; blue, 0 }  ,draw opacity=1 ][fill={rgb, 255:red, 255; green, 178; blue, 105 }  ,fill opacity=1 ][line width=0.75]  (247.33,74.5) .. controls (247.33,71.19) and (250.02,68.5) .. (253.33,68.5) .. controls (256.65,68.5) and (259.33,71.19) .. (259.33,74.5) .. controls (259.33,77.81) and (256.65,80.5) .. (253.33,80.5) .. controls (250.02,80.5) and (247.33,77.81) .. (247.33,74.5) -- cycle ;
\draw    (258.67,77.17) -- (281.67,95.17) ;

\draw (275.5,65.58) node   [align=left] {\begin{minipage}[lt]{22.89pt}\setlength\topsep{0pt}
{\fontfamily{pcr}\selectfont 0x00}
\end{minipage}};
\draw (307.83,91.25) node   [align=left] {\begin{minipage}[lt]{22.89pt}\setlength\topsep{0pt}
{\fontfamily{pcr}\selectfont 0x04}
\end{minipage}};
\draw (328.83,113.75) node   [align=left] {\begin{minipage}[lt]{22.89pt}\setlength\topsep{0pt}
{\fontfamily{pcr}\selectfont 0x08}
\end{minipage}};
\draw (266.33,136.75) node   [align=left] {\begin{minipage}[lt]{22.89pt}\setlength\topsep{0pt}
{\fontfamily{pcr}\selectfont 0x0C}
\end{minipage}};
\draw (354.15,144.6) node   [align=left] {\begin{minipage}[lt]{22.89pt}\setlength\topsep{0pt}
{\fontfamily{pcr}\selectfont 0x10}
\end{minipage}};
\draw (290.83,175.75) node   [align=left] {\begin{minipage}[lt]{22.89pt}\setlength\topsep{0pt}
{\fontfamily{pcr}\selectfont 0x14}
\end{minipage}};
\draw (371.83,175.25) node   [align=left] {\begin{minipage}[lt]{22.89pt}\setlength\topsep{0pt}
{\fontfamily{pcr}\selectfont 0x18}
\end{minipage}};

\end{tikzpicture}
    \caption{Coverage instrumentation using colors. Orange-colored nodes represent basic blocks executed during fuzzing, while uncolored nodes indicate unexplored paths.}
    \label{fig:ex_2}
\end{figure}

\subsubsection{Code Hooks for Targeted Fuzzing}\label{sec:code_hooks}
System-level firmware presents an enormous execution state space due to the presence of hardware peripherals, device drivers, interrupts, and privileged kernel operations. Exhaustively exploring all possible states is computationally infeasible. To address this challenge, SysFuSS integrates targeted instrumentation and adaptive exploration mechanisms to constrain and prioritize fuzzing within beneficial regions of the firmware.
A key element of this approach is a code hooking mechanism that directs the fuzzer toward specific code regions of interest. This selective focus is particularly valuable when analyzing large firmware images where exhaustive coverage is impractical or when investigating subsystems suspected of containing vulnerabilities. Each hook is defined by three parameters:

\begin{itemize}
\item \textbf{Starting address:} The entry point of the function or module to be fuzzed.
\item \textbf{Mutation size:} The size of the input buffer to be mutated while the hook is active, enabling fine-grained control over the fuzzing scope.
\item \textbf{Breakpoint:} The termination point where focused fuzzing concludes and normal execution resumes.
\end{itemize}

By using hooks to delimit specific execution regions, SysFuSS reduces the effective state space and allocates computational effort toward high-value targets, such as input parsers, protocol handlers, and cryptographic routines. This targeted approach improves both efficiency and coverage depth, allowing the fuzzer to explore complex firmware behavior without being overwhelmed by irrelevant peripheral interactions.




\subsubsection{Memory Hooks for Vulnerability Detection}\label{sec:memvul_detect}
Memory corruption vulnerabilities, including buffer overflows, use-after-free, and uninitialized memory reads, represent a significant class of firmware security flaws. To detect these issues during fuzzing, we implement a comprehensive memory monitoring system based on shadow memory techniques~\cite{gao2020fuzz,serebryany2012addresssanitizer}.
Our approach uses memory shadowing to maintain metadata about the state of each byte in the heap memory region. For every byte of heap memory, we maintain a corresponding shadow byte that tracks its current state. The shadow memory can be in one of four states:
\begin{enumerate}
\item \textbf{Unaddressable}: Memory that has not been allocated or has been freed. Any access to unaddressable memory indicates a serious bug, such as a buffer overflow or use-after-free vulnerability.
\item \textbf{Uninitialized}: Memory that has been allocated but not yet written. Reading from uninitialized memory can lead to information disclosure vulnerabilities or unpredictable program behavior.

\item \textbf{Defined}: Memory that has been both allocated and initialized with a value. This represents the normal, safe state for heap memory.

\item \textbf{Readable}: Memory that is allocated and may only be read but not written, typically used for const data or memory-mapped I/O regions.
\end{enumerate}
We implement memory hooks that intercept all heap memory operations, including \texttt{malloc}, \texttt{free}, \texttt{realloc}, and their variants. These hooks update the shadow memory state accordingly. For instance, when \texttt{malloc} allocates a new block, the corresponding shadow memory is marked as \textit{Uninitialized}. When the program writes to this memory, we transition it to the \textit{Defined} state.
Using our memory hooks in conjunction with shadow memory tracking, SysFuSS can identify a comprehensive range of memory vulnerabilities during firmware execution. Table~\ref{tab:memory_detectors} summarizes the memory detectors implemented in our framework and their underlying detection principles. Each detector leverages the shadow memory states to identify specific classes of memory safety violations. For example, buffer overflow detection monitors write to memory regions that are marked as unaddressable or readable. Similarly, use-after-free vulnerabilities are caught by detecting accesses to memory addresses marked as uninitialized or defined after they have been freed.

\begin{table}[ht]
\centering
\footnotesize
\caption{Detection of diverse memory vulnerabilities via shadow memory and hooking}
\label{tab:memory_detectors}
\begin{tabular}{@{}ll@{}}
\toprule
\textbf{Vulnerability} & \textbf{Detection Mechanism} \\ \midrule
Buffer Overflow & \begin{tabular}[c]{@{}l@{}}Detected when a write occurs beyond the\\ bounds of an allocated region.\end{tabular} \\ \midrule
Buffer Over-read & \begin{tabular}[c]{@{}l@{}}Detected when a read accesses memory\\ beyond the allocated region.\end{tabular} \\ \midrule
Buffer Underflow & \begin{tabular}[c]{@{}l@{}}Detected when a write accesses memory \\before the start of an allocated region.\end{tabular} \\ \midrule
Buffer Under-read & \begin{tabular}[c]{@{}l@{}}Detected when a read occurs before the base\\ address of an allocation.\end{tabular} \\ \midrule
Double Free & \begin{tabular}[c]{@{}l@{}}Detected when \texttt{free} is called on memory\\ already marked as \texttt{Unaddressable}.\end{tabular} \\ \midrule
Use-after-free & \begin{tabular}[c]{@{}l@{}}Detected when a read/write is attempted on\\ memory that has been freed.\end{tabular} \\ \midrule
Wild Free & \begin{tabular}[c]{@{}l@{}}Detected when free is called on an address \\that was never allocated.\end{tabular} \\ \midrule
Uninitialized Access & \begin{tabular}[c]{@{}l@{}}Detected when a read operation is performed \\on memory marked as \texttt{Uninitialized}.\end{tabular} \\ \midrule
Invalid read  & \begin{tabular}[c]{@{}l@{}}Detected when a read is attempted from an\\ aaddress not belonging to any valid allocation.\end{tabular} \\ \midrule
Invalid Write & \begin{tabular}[c]{@{}l@{}}Detected when a write operation is attempted\\ on memory marked as \texttt{Readable}.\end{tabular} \\\bottomrule
\end{tabular}
\end{table}

By integrating these memory detectors with our fuzzing framework, we can identify memory corruption bugs in real-time during test case execution. This is particularly valuable for firmware, where memory safety violations may not always result in immediate crashes but can lead to exploitable vulnerabilities. Our instrumentation provides detailed diagnostic information when violations are detected, including the exact memory address, operation type, and backtrace, facilitating rapid vulnerability triage and remediation.

\begin{figure}[!htb]
    \centering
    \small
    \tikzset{every picture/.style={line width=0.75pt}} 

\begin{tikzpicture}[x=0.75pt,y=0.75pt,yscale=-1,xscale=1]

\draw  [fill={rgb, 255:red, 255; green, 159; blue, 159 }  ,fill opacity=1 ] (80,54.6) -- (160,54.6) -- (160,80) -- (80,80) -- cycle ;
\draw  [fill={rgb, 255:red, 182; green, 213; blue, 255 }  ,fill opacity=1 ] (80,105) -- (160,105) -- (160,170) -- (80,170) -- cycle ;
\draw  [fill={rgb, 255:red, 174; green, 245; blue, 178 }  ,fill opacity=1 ] (80,170) -- (160,170) -- (160,195) -- (80,195) -- cycle ;
\draw  [fill={rgb, 255:red, 255; green, 247; blue, 134 }  ,fill opacity=1 ] (80,80) -- (160,80) -- (160,105) -- (80,105) -- cycle ;
\draw  [draw opacity=0][fill={rgb, 255:red, 255; green, 247; blue, 134 }  ,fill opacity=1 ] (205,99) .. controls (205,96.79) and (206.79,95) .. (209,95) -- (221,95) .. controls (223.21,95) and (225,96.79) .. (225,99) -- (225,111) .. controls (225,113.21) and (223.21,115) .. (221,115) -- (209,115) .. controls (206.79,115) and (205,113.21) .. (205,111) -- cycle ;
\draw  [draw opacity=0][fill={rgb, 255:red, 255; green, 159; blue, 159 }  ,fill opacity=1 ] (205,69) .. controls (205,66.79) and (206.79,65) .. (209,65) -- (221,65) .. controls (223.21,65) and (225,66.79) .. (225,69) -- (225,81) .. controls (225,83.21) and (223.21,85) .. (221,85) -- (209,85) .. controls (206.79,85) and (205,83.21) .. (205,81) -- cycle ;
\draw  [draw opacity=0][fill={rgb, 255:red, 182; green, 213; blue, 255 }  ,fill opacity=1 ] (205,129) .. controls (205,126.79) and (206.79,125) .. (209,125) -- (221,125) .. controls (223.21,125) and (225,126.79) .. (225,129) -- (225,141) .. controls (225,143.21) and (223.21,145) .. (221,145) -- (209,145) .. controls (206.79,145) and (205,143.21) .. (205,141) -- cycle ;
\draw  [draw opacity=0][fill={rgb, 255:red, 174; green, 245; blue, 178 }  ,fill opacity=1 ] (205,159) .. controls (205,156.79) and (206.79,155) .. (209,155) -- (221,155) .. controls (223.21,155) and (225,156.79) .. (225,159) -- (225,171) .. controls (225,173.21) and (223.21,175) .. (221,175) -- (209,175) .. controls (206.79,175) and (205,173.21) .. (205,171) -- cycle ;

\draw (60,195) node   [align=left] {\begin{minipage}[lt]{27.2pt}\setlength\topsep{0pt}
{\scriptsize {\fontfamily{pcr}\selectfont 0x1000}}
\end{minipage}};
\draw (60,170) node   [align=left] {\begin{minipage}[lt]{27.2pt}\setlength\topsep{0pt}
{\scriptsize {\fontfamily{pcr}\selectfont 0x1008}}
\end{minipage}};
\draw (60,105) node   [align=left] {\begin{minipage}[lt]{27.2pt}\setlength\topsep{0pt}
{\scriptsize {\fontfamily{pcr}\selectfont 0x1018}}
\end{minipage}};
\draw (60,80) node   [align=left] {\begin{minipage}[lt]{27.2pt}\setlength\topsep{0pt}
{\scriptsize {\fontfamily{pcr}\selectfont 0x101C}}
\end{minipage}};
\draw (60,55) node   [align=left] {\begin{minipage}[lt]{27.2pt}\setlength\topsep{0pt}
{\scriptsize {\fontfamily{pcr}\selectfont 0x1020}}
\end{minipage}};
\draw (120,182.5) node   [align=left] {\begin{minipage}[lt]{54.4pt}\setlength\topsep{0pt}
\begin{center}
const data
\end{center}

\end{minipage}};
\draw (120,137.3) node   [align=left] {\begin{minipage}[lt]{54.4pt}\setlength\topsep{0pt}
\begin{center}
byte array
\end{center}

\end{minipage}};
\draw (120,92.5) node   [align=left] {\begin{minipage}[lt]{54.4pt}\setlength\topsep{0pt}
\begin{center}
(Allocated)
\end{center}

\end{minipage}};
\draw (120,67.5) node   [align=left] {\begin{minipage}[lt]{54.4pt}\setlength\topsep{0pt}
\begin{center}
(Freed)
\end{center}

\end{minipage}};
\draw (282.5,105) node   [align=left] {\begin{minipage}[lt]{71.4pt}\setlength\topsep{0pt}
Unaddressable
\end{minipage}};
\draw (282.5,75) node   [align=left] {\begin{minipage}[lt]{71.4pt}\setlength\topsep{0pt}
Uninitialized
\end{minipage}};
\draw (282.5,135) node   [align=left] {\begin{minipage}[lt]{71.4pt}\setlength\topsep{0pt}
Defined
\end{minipage}};
\draw (282.5,165) node   [align=left] {\begin{minipage}[lt]{71.4pt}\setlength\topsep{0pt}
Readable
\end{minipage}};

\end{tikzpicture}
    \caption{Sample memory layout.}
    \label{fig:ex_mem_hook}
\end{figure}
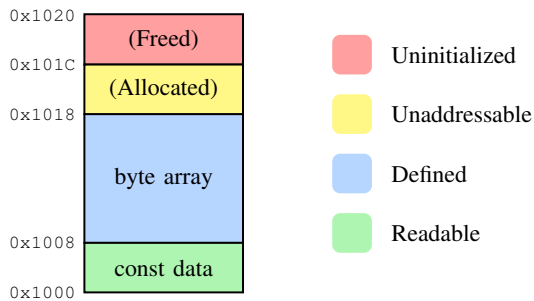

{
\vspace{0.05in}
\noindent \underline{\textbf{Example 3} (Memory Hooks):} Consider the memory layout shown in Figure \ref{fig:ex_mem_hook}, where firmware has allocated a constant data region from 0x1000 to 0x1007 and a byte array of 16 bytes from 0x1008 to 0x1017. Above the byte array lies a freed memory region from 0x1018 to 0x101B and an allocated but uninitialized region from 0x101C to 0x101F. Suppose the firmware attempts to write 20 bytes into the 16-byte array. The memory hooks intercept each write operation and check the corresponding shadow memory state. The first 16 bytes (0x1008 to 0x1017) succeed because shadow memory indicates a ``Defined" state (\textcolor[HTML]{B6D5FF}{$\blacksquare$}). However, when the 17th byte attempts to write to address 0x1018, the hook queries shadow memory and discovers an ``Unaddressable" state (\textcolor[HTML]{FFF786}{$\blacksquare$}), indicating freed memory beyond the buffer bounds, immediately triggering a buffer overflow alert. In another scenario, if firmware attempts to modify the const data region by writing to address 0x1000, the memory hook intercepts the write and queries shadow memory, which returns a ``Readable" state (\textcolor[HTML]{AEF5B2}{$\blacksquare$}) indicating read-only memory. The hook immediately flags this as an invalid write operation, preventing corruption of immutable data. Similarly, if firmware reads from the uninitialized region at 0x101C without prior initialization, the shadow memory state ``Uninitialized" (\textcolor[HTML]{FF9F9F}{$\blacksquare$}) triggers an uninitialized access violation. Through this shadow memory tracking mechanism, memory hooks can detect all vulnerability scenarios listed in Table \ref{tab:memory_detectors}.
\exampleEnd
}

\subsection{Fuzzing and Plateau Detection} \label{sec:plateau_detection}

After instrumenting the firmware, we initialize the fuzzing engine with a set of seed inputs, valid test cases that exercise the target firmware without causing crashes. As fuzzing progresses, we continuously monitor branch coverage by tracking which edges in the previously constructed CFG have been explored. Initially, the fuzzer rapidly discovers new code paths, but this progress eventually slows as the easily reachable portions of the code are exhausted. The critical challenge is identifying the precise moment when fuzzing has reached diminishing returns, indicating that symbolic execution should be invoked to break through hard-to-reach regions. To address this, SysFuSS continues fuzzing until a coverage plateau is detected, after which it generates the necessary path constraints for symbolic execution, as described in lines 6–8 of Algorithm~\ref{alg:sysfuss}.

\subsubsection{The Challenge of Plateau Detection}
Detecting a coverage plateau is not straightforward because coverage metrics exhibit natural fluctuations even when meaningful progress has stalled. A naive approach that triggers symbolic execution after any brief period of stagnation would be too aggressive, wasting computational resources on premature transitions. Conversely, waiting too long after the plateau begins wastes fuzzing cycles that could be better spent on symbolic analysis. We need a robust detection mechanism that distinguishes between temporary stagnation and genuine coverage plateaus while remaining responsive enough to trigger symbolic execution in a timely manner.
\subsubsection{Formalization of Coverage Plateau Detection}
To formalize the coverage plateau detection, we define a cumulative coverage function that captures the evolution of code coverage over time:
\begin{equation}
C(t) = \# \text{ of unique branches covered at iteration } i
\label{eq:coverage_function}
\end{equation}
Here, $i$ represents the number of fuzzing iterations since the start of the fuzzing campaign. The function $C(i)$ is monotonically non-decreasing, as new branches may be discovered but previously covered branches remain marked as explored.

To detect plateaus, we employ a sliding window approach that examines coverage trends over recent history rather than relying on instantaneous measurements. We define a window size $W$ that specifies the number of recent iterations to consider when evaluating coverage progress. The choice of $W$ involves a trade-off: smaller values make the detector more responsive to changes but more susceptible to noise, while larger values provide stability but delay plateau detection. In practice, $W$ is a tunable parameter that can be adjusted based on the firmware's complexity and the fuzzer's typical exploration rate.
Within this sliding window, we compute the absolute coverage increment:
\begin{equation}
\Delta C(i, W) = C(i) - C(i - W)
\label{eq:coverage_increment}
\end{equation}
This quantity represents the total number of new branches discovered during the most recent $W$ iterations. To normalize this measure and make it independent of window size, we calculate the average coverage improvement per iteration:

\begin{equation}
\overline{\Delta C}(i, W) =\frac{\Delta C(i, W)}{W}
\label{eq:avg_coverage_increment}
\end{equation}
The normalized metric $\overline{\Delta C}(i, W)$ represents the rate of coverage growth, expressed as the average number of new branches discovered per iteration within the sliding window. This rate-based measure allows us to establish a meaningful threshold that is independent of the absolute coverage achieved.

\subsubsection{Detecting the Coverage Plateau}
We define a coverage increment threshold $\epsilon \in \mathbb{R}^+$ that represents the minimum acceptable rate of coverage growth. When the average coverage improvement falls below this threshold, we conclude that fuzzing has reached a point of diminishing returns. The threshold $\epsilon$ is typically set to a small positive value based on empirical observations of the fuzzer's behavior.
Using this threshold, we define a binary plateau detection function:
\begin{equation}
Detect(i) = \begin{cases}
1  & \text{if } \overline{\Delta C}(i, W) < \epsilon \\
0 & \text{otherwise}
\end{cases}
\label{eq:plateau_function}
\end{equation}
When $Detect(i) = 1$, the system signals that a coverage plateau has been detected at iteration $i$, triggering a transition to symbolic execution. When $Detect(i)=0$, fuzzing continues normally as coverage is still improving at an acceptable rate. Specifically, the plateau detection logic is implemented in line 8 of Algorithm~\ref{alg:sysfuss}.

\subsection{Constraint Generation}
\label{sec:source_dest_extraction}
Upon detecting a coverage plateau, the framework must transition to symbolic execution. However, directly applying symbolic execution to the entire firmware is computationally expensive. Instead, we strategically target specific unexplored regions by identifying constraint pairs. These pairs of program locations where a covered branch (source) leads to an uncovered branch (destination). These pairs represent the frontier between explored and unexplored code, making them ideal candidates for symbolic analysis. This process corresponds to the lines 9--11 in Algorithm~\ref{alg:sysfuss}.

\subsubsection{Identifying Unexplored Regions}
We extract constraint pairs by analyzing the instrumented CFG in conjunction with the coverage information collected during fuzzing. For each edge $e = (u, v)$ in the CFG where $u$ has been covered but $v$ remains unexplored, we create a constraint pair $(s,d)$ where $s=u$ represents the source node (last covered location) and $d=v$ represents the destination node (unexplored target). Formally, the set of candidate constraint pairs is defined as:
\begin{equation}
CP = {(s_i, d_i) \mid s_i \in V_{c}, d_i \in V_{c}', (s_i, d_i) \in E}
\label{eq:sd_set}
\end{equation}
where $V_{c}$ denotes the set of covered nodes, $V_{c}'$ denotes the set of uncovered nodes, and $E$ represents the edges in the CFG. In practice, complex firmware designs may have thousands of such pairs, making it essential to prioritize which regions to explore first.

\begin{figure}[!htb]
    \centering
    \small
    \tikzset{every picture/.style={line width=0.75pt}} 

\begin{tikzpicture}[x=0.75pt,y=0.75pt,yscale=-1,xscale=1]

\draw  [color={rgb, 255:red, 0; green, 0; blue, 0 }  ,draw opacity=1 ][fill={rgb, 255:red, 255; green, 178; blue, 105 }  ,fill opacity=1 ][line width=0.75]  (280,88.5) .. controls (280,85.19) and (282.69,82.5) .. (286,82.5) .. controls (289.31,82.5) and (292,85.19) .. (292,88.5) .. controls (292,91.81) and (289.31,94.5) .. (286,94.5) .. controls (282.69,94.5) and (280,91.81) .. (280,88.5) -- cycle ;
\draw  [color={rgb, 255:red, 0; green, 0; blue, 0 }  ,draw opacity=1 ][fill={rgb, 255:red, 255; green, 178; blue, 105 }  ,fill opacity=1 ][line width=0.75]  (308,115.83) .. controls (308,112.52) and (310.69,109.83) .. (314,109.83) .. controls (317.31,109.83) and (320,112.52) .. (320,115.83) .. controls (320,119.15) and (317.31,121.83) .. (314,121.83) .. controls (310.69,121.83) and (308,119.15) .. (308,115.83) -- cycle ;
\draw  [color={rgb, 255:red, 0; green, 0; blue, 0 }  ,draw opacity=1 ][fill={rgb, 255:red, 255; green, 178; blue, 105 }  ,fill opacity=1 ][line width=0.75]  (330.63,145.19) .. controls (330.63,141.77) and (333.4,139) .. (336.81,139) .. controls (340.23,139) and (343,141.77) .. (343,145.19) .. controls (343,148.6) and (340.23,151.38) .. (336.81,151.38) .. controls (333.4,151.38) and (330.63,148.6) .. (330.63,145.19) -- cycle ;
\draw    (290.5,91.83) -- (309.5,111.17) ;
\draw    (308.83,119.83) -- (290.5,136.83) ;
\draw    (317.5,120.75) -- (333.5,140.25) ;
\draw    (333.33,150) -- (316.33,176.33) ;
\draw    (357.67,175.33) -- (340.67,150) ;
\draw  [color={rgb, 255:red, 0; green, 0; blue, 0 }  ,draw opacity=1 ][line width=0.75]  (280,141.5) .. controls (280,138.19) and (282.69,135.5) .. (286,135.5) .. controls (289.31,135.5) and (292,138.19) .. (292,141.5) .. controls (292,144.81) and (289.31,147.5) .. (286,147.5) .. controls (282.69,147.5) and (280,144.81) .. (280,141.5) -- cycle ;
\draw  [color={rgb, 255:red, 0; green, 0; blue, 0 }  ,draw opacity=1 ][line width=0.75]  (307,181) .. controls (307,177.69) and (309.69,175) .. (313,175) .. controls (316.31,175) and (319,177.69) .. (319,181) .. controls (319,184.31) and (316.31,187) .. (313,187) .. controls (309.69,187) and (307,184.31) .. (307,181) -- cycle ;
\draw  [color={rgb, 255:red, 0; green, 0; blue, 0 }  ,draw opacity=1 ][fill={rgb, 255:red, 255; green, 178; blue, 105 }  ,fill opacity=1 ][line width=0.75]  (355,181) .. controls (355,177.69) and (357.69,175) .. (361,175) .. controls (364.31,175) and (367,177.69) .. (367,181) .. controls (367,184.31) and (364.31,187) .. (361,187) .. controls (357.69,187) and (355,184.31) .. (355,181) -- cycle ;
\draw  [color={rgb, 255:red, 0; green, 0; blue, 0 }  ,draw opacity=1 ][fill={rgb, 255:red, 255; green, 178; blue, 105 }  ,fill opacity=1 ][line width=0.75]  (251.33,63.5) .. controls (251.33,60.19) and (254.02,57.5) .. (257.33,57.5) .. controls (260.65,57.5) and (263.33,60.19) .. (263.33,63.5) .. controls (263.33,66.81) and (260.65,69.5) .. (257.33,69.5) .. controls (254.02,69.5) and (251.33,66.81) .. (251.33,63.5) -- cycle ;
\draw    (261.5,67.5) -- (281.17,84.17) ;
\draw    (252,67) -- (230,84.33) ;
\draw  [color={rgb, 255:red, 0; green, 0; blue, 0 }  ,draw opacity=1 ][fill={rgb, 255:red, 255; green, 178; blue, 105 }  ,fill opacity=1 ][line width=0.75]  (219.5,89) .. controls (219.5,85.69) and (222.19,83) .. (225.5,83) .. controls (228.81,83) and (231.5,85.69) .. (231.5,89) .. controls (231.5,92.31) and (228.81,95) .. (225.5,95) .. controls (222.19,95) and (219.5,92.31) .. (219.5,89) -- cycle ;
\draw    (221.33,93.33) -- (206.5,111.5) ;
\draw  [color={rgb, 255:red, 0; green, 0; blue, 0 }  ,draw opacity=1 ][line width=0.75]  (197,116.33) .. controls (197,113.02) and (199.69,110.33) .. (203,110.33) .. controls (206.31,110.33) and (209,113.02) .. (209,116.33) .. controls (209,119.65) and (206.31,122.33) .. (203,122.33) .. controls (199.69,122.33) and (197,119.65) .. (197,116.33) -- cycle ;
\draw    (229.5,93) -- (245,111) ;
\draw  [color={rgb, 255:red, 0; green, 0; blue, 0 }  ,draw opacity=1 ][line width=0.75]  (242.5,116.43) .. controls (242.5,113.07) and (245.16,110.36) .. (248.43,110.36) .. controls (251.71,110.36) and (254.36,113.07) .. (254.36,116.43) .. controls (254.36,119.78) and (251.71,122.5) .. (248.43,122.5) .. controls (245.16,122.5) and (242.5,119.78) .. (242.5,116.43) -- cycle ;
\draw    (198.83,121.33) -- (184,139.5) ;
\draw  [color={rgb, 255:red, 0; green, 0; blue, 0 }  ,draw opacity=1 ][line width=0.75]  (174.5,144.33) .. controls (174.5,141.02) and (177.19,138.33) .. (180.5,138.33) .. controls (183.81,138.33) and (186.5,141.02) .. (186.5,144.33) .. controls (186.5,147.65) and (183.81,150.33) .. (180.5,150.33) .. controls (177.19,150.33) and (174.5,147.65) .. (174.5,144.33) -- cycle ;
\draw    (207,121) -- (222.5,139) ;
\draw  [color={rgb, 255:red, 0; green, 0; blue, 0 }  ,draw opacity=1 ][line width=0.75]  (220,144.43) .. controls (220,141.07) and (222.66,138.36) .. (225.93,138.36) .. controls (229.21,138.36) and (231.86,141.07) .. (231.86,144.43) .. controls (231.86,147.78) and (229.21,150.5) .. (225.93,150.5) .. controls (222.66,150.5) and (220,147.78) .. (220,144.43) -- cycle ;
\draw   (187.29,216.14) -- (236,216.14) -- (236,240) -- (187.29,240) -- cycle ;
\draw   (236,216.14) -- (283.29,216.14) -- (283.29,240) -- (236,240) -- cycle ;
\draw   (283.29,216.14) -- (331.29,216.14) -- (331.29,240) -- (283.29,240) -- cycle ;
\draw   (331.29,216) -- (380,216) -- (380,240) -- (331.29,240) -- cycle ;
\draw   (160,52.5) .. controls (160,45.6) and (165.6,40) .. (172.5,40) -- (367.5,40) .. controls (374.4,40) and (380,45.6) .. (380,52.5) -- (380,190.25) .. controls (380,197.15) and (374.4,202.75) .. (367.5,202.75) -- (172.5,202.75) .. controls (165.6,202.75) and (160,197.15) .. (160,190.25) -- cycle ;

\draw (212.25,78) node   [align=left] {\begin{minipage}[lt]{14.62pt}\setlength\topsep{0pt}
$\displaystyle s_{1}$
\end{minipage}};
\draw (330.75,103) node   [align=left] {\begin{minipage}[lt]{14.62pt}\setlength\topsep{0pt}
$\displaystyle s_{2}$
\end{minipage}};
\draw (351.42,134.17) node   [align=left] {\begin{minipage}[lt]{14.62pt}\setlength\topsep{0pt}
$\displaystyle s_{3}$
\end{minipage}};
\draw (194.75,152.33) node   [align=left] {\begin{minipage}[lt]{14.62pt}\setlength\topsep{0pt}
$\displaystyle d_{1}$
\end{minipage}};
\draw (265.75,110.83) node   [align=left] {\begin{minipage}[lt]{14.62pt}\setlength\topsep{0pt}
$\displaystyle d_{2}$
\end{minipage}};
\draw (275.25,148.33) node   [align=left] {\begin{minipage}[lt]{14.62pt}\setlength\topsep{0pt}
$\displaystyle d_{3}$
\end{minipage}};
\draw (300.25,181.33) node   [align=left] {\begin{minipage}[lt]{14.62pt}\setlength\topsep{0pt}
$\displaystyle d_{4}$
\end{minipage}};
\draw (256.75,46.33) node   [align=left] {\begin{minipage}[lt]{14.62pt}\setlength\topsep{0pt}
$\displaystyle v_{0}$
\end{minipage}};
\draw (211.64,228.07) node   [align=left] {\begin{minipage}[lt]{33.13pt}\setlength\topsep{0pt}
$\displaystyle ( s_{1} ,\ d_{1})$
\end{minipage}};
\draw (259.79,228) node   [align=left] {\begin{minipage}[lt]{32.15pt}\setlength\topsep{0pt}
$\displaystyle ( s_{1} ,\ d_{2})$
\end{minipage}};
\draw (355.64,228) node   [align=left] {\begin{minipage}[lt]{33.13pt}\setlength\topsep{0pt}
$\displaystyle ( s_{3} ,\ d_{4})$
\end{minipage}};
\draw (168.21,227.71) node   [align=left] {\begin{minipage}[lt]{33.13pt}\setlength\topsep{0pt}
$\displaystyle CP=$
\end{minipage}};
\draw (306.93,228.07) node   [align=left] {\begin{minipage}[lt]{32.15pt}\setlength\topsep{0pt}
$\displaystyle ( s_{2} ,\ d_{3})$
\end{minipage}};

\end{tikzpicture}
    \caption{An example of constraint pair (CP) generation}
    \label{fig:sd_extraction}
\end{figure}

{
\vspace{0.05in}
\noindent \underline{\textbf{Example 4} (Constraint Extraction):} Figure \ref{fig:sd_extraction} illustrates the CFG of the firmware after several fuzzing iterations. The figure demonstrates how constraint pairs are systematically identified at the boundary between explored and unexplored code. Starting from the entry node $v_0$, we perform a depth-first traversal of the CFG. Whenever the traversal encounters an edge leading from a covered node to an uncovered node, we record the covered node as a source point. For each source, we identify a corresponding destination by selecting a leaf node (or deep node) within the unexplored subtree rooted at the uncovered child. 
\exampleEnd
}

\subsubsection{Prioritization Strategy}
To manage the potentially large set of constraint pairs efficiently, we employ a priority queue that orders pairs based on their distance from the CFG entry point. This breadth-first exploration strategy prioritizes shallow unexplored regions over deep ones, enabling symbolic execution to uncover large contiguous areas of new code. Once these regions are added to the fuzzing corpus, subsequent fuzzing campaigns can explore them through mutation-based techniques. This approach minimizes the number of expensive symbolic execution invocations while maximizing coverage gains.
We define a scoring function $\sigma(s,d) \rightarrow \mathbb{N}$ that assigns each constraint pair a priority score based on the shortest path distance from the CFG entry node $v_0$ to the source node:

\begin{equation}
\sigma(s_i, d_i) = \text{dist}(v_0, s_i)
\label{eq:score_function}
\end{equation}
where $\text{dist}(v_0, s_i)$ computes the minimum number of edges in any path from the entry point $v_0$ to source node
$s_i$. Pairs with smaller scores are prioritized, ensuring that the symbolic execution engine targets shallow unexplored branches before deeper ones.

The priority queue $\mathcal{Q}$ is maintained as a min-heap ordered by the scoring function:

\begin{equation}
\mathcal{Q} = \text{MinHeap}(CP, \sigma)
\label{eq:priority_queue}
\end{equation}
To maintain computational feasibility, we impose a maximum queue size $L$ that limits the number of constraint pairs considered in each symbolic execution phase: $|\mathcal{Q}| \leq L$
When $|\mathcal{Q}| > L$, we retain only the top $L$ pairs with the smallest scores, effectively pruning deep unexplored regions that would be expensive to reach symbolically. The resulting priority queue $\mathcal{Q}$ is then passed to the symbolic execution engine to generate concrete test inputs that reach the targeted destination nodes.


\vspace{0.05in}
\noindent \underline{{\textbf{Example 5} (Constraint Prioritization)}}: To illustrate the construction and prioritization of the priority queue, we present a concrete example using the CFG in Figure~\ref{fig:sd_extraction}. Consider two constraint pairs: $(s_1, d_1)$ and $(s_3, d_4)$, as shown in the diagram. To maximize coverage of unexplored regions, we must prioritize these branches appropriately. We apply the scoring function $\sigma(s_i, d_i) = \text{dist}(v_0, s_i)$ from Equation~\ref{eq:score_function} to each pair. Since $s_1$ is located 2 edges from the entry point $v_0$, we have $\sigma(s_1, d_1) = 2$. Similarly, $s_3$ is 4 edges from the entry point, yielding $\sigma(s_3, d_4) = 4$. The priority queue $\mathcal{Q}$ orders these pairs by ascending score, placing $(s_1, d_1)$ ahead of $(s_3, d_4)$ due to its proximity to the entry point. 


\subsection{Selective Symbolic Execution}
\label{sec:symbolic_execution}
Once the priority queue of constraint pairs is constructed, we employ symbolic execution to generate concrete inputs that can reach the unexplored destination nodes. Unlike traditional symbolic execution that begins from the program entry point, our approach leverages the concrete execution state at the source node to initialize symbolic execution at an intermediate program location. This targeted strategy significantly reduces the path explosion problem and improves the efficiency of constraint solving. This entire process is illustrated by the lines 12--20 in Algorithm~\ref{alg:sysfuss}.

\subsubsection{Initialization from Intermediate States}
For each constraint pair $(s_i, d_i) \in \mathcal{Q}$, we configure the symbolic execution engine to begin analysis at the source node $s_i$ rather than the firmware entry point $v_0$. To achieve this, we extract the concrete program state $\Sigma_i$ that was recorded when the fuzzer last executed node $s_i$ during the fuzzing phase. This state includes:

\begin{equation}
\Sigma_i = \langle \text{PC}_i, \text{Regs}_i, \text{Mem}_i \rangle
\label{eq:concrete_state}
\end{equation}
where $\text{PC}_i$ denotes the program counter value at $s_i$, $\text{Regs}_i$ represents the register file state, and $\text{Mem}_i$ captures the relevant memory contents. We restore the symbolic execution engine to this concrete state $\Sigma_i$, effectively resuming execution from the last covered node $s_i$ by the fuzzer.

Next, we identify the portion of the input that influences the branch condition at $s_i$ leading to $d_i$. Rather than marking the entire input as symbolic (which would create an intractable constraint system), we selectively mark only the relevant input bytes ($I_s$) as symbolic variables:

\begin{equation}
I_s = {x_j \mid x_j \text{ influences path condition at } s_i}
\label{eq:symbolic_input}
\end{equation}
This selective symbolization is achieved through dynamic taint analysis, which tracks how input bytes propagate through the computation and affect the branch condition. By constraining the symbolic space to only the relevant variables, we significantly reduce the complexity of the symbolic execution task.
\subsubsection{Path Solving}
With the symbolic execution engine initialized at state $\Sigma_i$ and the relevant input bytes marked as symbolic, we configure the destination target as node $d_i$. The symbolic execution engine then explores execution paths starting from $s_i$, accumulating path constraints along each explored path. For the path leading from $s_i$ to $d_i$, the engine constructs a path constraint $\phi_{s_i \rightarrow d_i}$ that represents the logical conditions on symbolic inputs required to reach $d_i$:

\begin{equation}
\phi_{s_i \rightarrow d_i} = \bigwedge_{k=1}^{m} c_k
\label{eq:path_constraint}
\end{equation}
where $c_k$ denotes individual branch conditions encountered along the path from $s_i$ to $d_i$, and $m$ is the number of conditional branches on this path. The symbolic execution engine invokes an SMT solver to determine the satisfiability of $\phi_{s_i \rightarrow d_i}$.
\begin{equation}
\text{SAT}(\phi_{s_i \rightarrow d_i}) \implies \exists I_c : \phi_{s_i \rightarrow d_i}(I_c) = \text{true}
\label{eq:sat_solving}
\end{equation}

\begin{figure}[!htb]
    \centering
    \small
    \tikzset{every picture/.style={line width=0.75pt}} 

\begin{tikzpicture}[x=0.75pt,y=0.75pt,yscale=-1,xscale=1]

\draw  [color={rgb, 255:red, 0; green, 0; blue, 0 }  ,draw opacity=1 ][fill={rgb, 255:red, 255; green, 178; blue, 105 }  ,fill opacity=1 ][line width=0.75]  (249.42,103.54) .. controls (249.42,100.23) and (252.1,97.54) .. (255.42,97.54) .. controls (258.73,97.54) and (261.42,100.23) .. (261.42,103.54) .. controls (261.42,106.86) and (258.73,109.54) .. (255.42,109.54) .. controls (252.1,109.54) and (249.42,106.86) .. (249.42,103.54) -- cycle ;
\draw  [color={rgb, 255:red, 0; green, 0; blue, 0 }  ,draw opacity=1 ][fill={rgb, 255:red, 255; green, 178; blue, 105 }  ,fill opacity=1 ][line width=0.75]  (193.58,166.17) .. controls (193.58,162.85) and (196.27,160.17) .. (199.58,160.17) .. controls (202.9,160.17) and (205.58,162.85) .. (205.58,166.17) .. controls (205.58,169.48) and (202.9,172.17) .. (199.58,172.17) .. controls (196.27,172.17) and (193.58,169.48) .. (193.58,166.17) -- cycle ;
\draw  [color={rgb, 255:red, 0; green, 0; blue, 0 }  ,draw opacity=1 ][line width=0.75]  (242.71,162.38) .. controls (242.71,158.96) and (245.48,156.19) .. (248.9,156.19) .. controls (252.31,156.19) and (255.08,158.96) .. (255.08,162.38) .. controls (255.08,165.8) and (252.31,168.57) .. (248.9,168.57) .. controls (245.48,168.57) and (242.71,165.8) .. (242.71,162.38) -- cycle ;
\draw  [color={rgb, 255:red, 0; green, 0; blue, 0 }  ,draw opacity=1 ][fill={rgb, 255:red, 255; green, 178; blue, 105 }  ,fill opacity=1 ][line width=0.75]  (219.67,133.92) .. controls (219.67,130.6) and (222.35,127.92) .. (225.67,127.92) .. controls (228.98,127.92) and (231.67,130.6) .. (231.67,133.92) .. controls (231.67,137.23) and (228.98,139.92) .. (225.67,139.92) .. controls (222.35,139.92) and (219.67,137.23) .. (219.67,133.92) -- cycle ;
\draw  [dash pattern={on 0.84pt off 2.51pt}]  (251.25,108.38) -- (250.75,107.13) -- (230.75,129.63) ;
\draw    (221.83,139.04) -- (202.92,161.5) ;
\draw  [dash pattern={on 0.84pt off 2.51pt}]  (229.17,138.79) -- (244.2,158.6) ;
\draw    (244.83,167.19) -- (227.83,193.19) ;
\draw    (271.33,192.69) -- (253.08,167.19) ;
\draw  [color={rgb, 255:red, 0; green, 0; blue, 0 }  ,draw opacity=1 ][line width=0.75]  (219.08,198.57) .. controls (219.08,195.25) and (221.77,192.57) .. (225.08,192.57) .. controls (228.4,192.57) and (231.08,195.25) .. (231.08,198.57) .. controls (231.08,201.88) and (228.4,204.57) .. (225.08,204.57) .. controls (221.77,204.57) and (219.08,201.88) .. (219.08,198.57) -- cycle ;
\draw  [color={rgb, 255:red, 0; green, 0; blue, 0 }  ,draw opacity=1 ][line width=0.75]  (267.08,198.57) .. controls (267.08,195.25) and (269.77,192.57) .. (273.08,192.57) .. controls (276.4,192.57) and (279.08,195.25) .. (279.08,198.57) .. controls (279.08,201.88) and (276.4,204.57) .. (273.08,204.57) .. controls (269.77,204.57) and (267.08,201.88) .. (267.08,198.57) -- cycle ;
\draw  [color={rgb, 255:red, 0; green, 0; blue, 0 }  ,draw opacity=1 ][fill={rgb, 255:red, 255; green, 178; blue, 105 }  ,fill opacity=1 ][line width=0.75]  (289.42,65.88) .. controls (289.42,62.56) and (292.1,59.88) .. (295.42,59.88) .. controls (298.73,59.88) and (301.42,62.56) .. (301.42,65.88) .. controls (301.42,69.19) and (298.73,71.88) .. (295.42,71.88) .. controls (292.1,71.88) and (289.42,69.19) .. (289.42,65.88) -- cycle ;
\draw    (291.33,69.83) -- (260.42,99.13) ;
\draw [color={rgb, 255:red, 0; green, 0; blue, 0 }  ,draw opacity=1 ][line width=1.5]  [dash pattern={on 5.63pt off 4.5pt}]  (287.75,57.63) .. controls (255.42,73.79) and (218.54,116.81) .. (195.06,147.37) ;
\draw [shift={(192.92,150.17)}, rotate = 307.27] [fill={rgb, 255:red, 0; green, 0; blue, 0 }  ,fill opacity=1 ][line width=0.08]  [draw opacity=0] (13.4,-6.43) -- (0,0) -- (13.4,6.44) -- (8.9,0) -- cycle    ;
\draw [color={rgb, 255:red, 163; green, 79; blue, 183 }  ,draw opacity=1 ][line width=1.5]    (234.2,133) .. controls (238.02,132.09) and (244.81,132.01) .. (243.9,138.56) .. controls (243,145.1) and (244.09,147.65) .. (250.63,145.47) .. controls (257.18,143.28) and (259.72,146.74) .. (257.9,151.28) .. controls (256.09,155.83) and (257.36,160.38) .. (264.81,160.56) .. controls (272.27,160.74) and (259.72,172.38) .. (267.9,173.1) .. controls (275.47,173.78) and (274.64,169.56) .. (275.7,183.72) ;
\draw [shift={(276,187.54)}, rotate = 265.28] [fill={rgb, 255:red, 163; green, 79; blue, 183 }  ,fill opacity=1 ][line width=0.08]  [draw opacity=0] (13.4,-6.43) -- (0,0) -- (13.4,6.44) -- (8.9,0) -- cycle    ;
\draw [color={rgb, 255:red, 0; green, 0; blue, 0 }  ,draw opacity=1 ][line width=1.5]  [dash pattern={on 5.63pt off 4.5pt}]  (356.17,174.87) .. controls (368,174.87) and (369.61,175.14) .. (381.61,175.19) ;
\draw [shift={(385.5,175.2)}, rotate = 180] [fill={rgb, 255:red, 0; green, 0; blue, 0 }  ,fill opacity=1 ][line width=0.08]  [draw opacity=0] (13.4,-6.43) -- (0,0) -- (13.4,6.44) -- (8.9,0) -- cycle    ;
\draw [color={rgb, 255:red, 163; green, 79; blue, 183 }  ,draw opacity=1 ][line width=1.5]    (356.43,199.6) .. controls (361.03,195) and (363.43,193.4) .. (364.83,197) .. controls (366.23,200.6) and (367.23,204.8) .. (371.03,201) .. controls (373.94,198.09) and (377.18,198.05) .. (381.24,197.97) ;
\draw [shift={(385.23,197.8)}, rotate = 175.33] [fill={rgb, 255:red, 163; green, 79; blue, 183 }  ,fill opacity=1 ][line width=0.08]  [draw opacity=0] (11.07,-5.32) -- (0,0) -- (11.07,5.32) -- (7.35,0) -- cycle    ;
\draw  [color={rgb, 255:red, 255; green, 209; blue, 143 }  ,draw opacity=1 ][line width=2.25]  (175,80.65) .. controls (175,62.2) and (189.95,47.25) .. (208.4,47.25) -- (308.6,47.25) .. controls (327.05,47.25) and (342,62.2) .. (342,80.65) -- (342,186.6) .. controls (342,205.05) and (327.05,220) .. (308.6,220) -- (208.4,220) .. controls (189.95,220) and (175,205.05) .. (175,186.6) -- cycle ;
\draw    (317.83,95.69) -- (299.58,70.19) ;
\draw  [color={rgb, 255:red, 0; green, 0; blue, 0 }  ,draw opacity=1 ][line width=0.75]  (313.58,101.57) .. controls (313.58,98.25) and (316.27,95.57) .. (319.58,95.57) .. controls (322.9,95.57) and (325.58,98.25) .. (325.58,101.57) .. controls (325.58,104.88) and (322.9,107.57) .. (319.58,107.57) .. controls (316.27,107.57) and (313.58,104.88) .. (313.58,101.57) -- cycle ;
\draw  [color={rgb, 255:red, 190; green, 119; blue, 255 }  ,draw opacity=1 ][line width=2.25]  (219.44,123.09) .. controls (230.23,114.42) and (255.15,127.53) .. (275.09,152.36) .. controls (295.04,177.19) and (302.45,204.35) .. (291.66,213.01) .. controls (280.87,221.68) and (255.95,208.58) .. (236.01,183.75) .. controls (216.06,158.92) and (208.65,131.76) .. (219.44,123.09) -- cycle ;
\draw    (369.67,73.33) -- (344.9,79.88) ;
\draw [shift={(342,80.65)}, rotate = 345.19] [fill={rgb, 255:red, 0; green, 0; blue, 0 }  ][line width=0.08]  [draw opacity=0] (10.72,-5.15) -- (0,0) -- (10.72,5.15) -- (7.12,0) -- cycle    ;
\draw    (369,122.67) -- (294.17,172.34) ;
\draw [shift={(291.67,174)}, rotate = 326.42] [fill={rgb, 255:red, 0; green, 0; blue, 0 }  ][line width=0.08]  [draw opacity=0] (10.72,-5.15) -- (0,0) -- (10.72,5.15) -- (7.12,0) -- cycle    ;

\draw (225.58,145.83) node  [font=\scriptsize] [align=left] {\begin{minipage}[lt]{8.67pt}\setlength\topsep{0pt}
$\displaystyle s_{i}$
\end{minipage}};
\draw (284.58,202.57) node  [font=\scriptsize] [align=left] {\begin{minipage}[lt]{8.67pt}\setlength\topsep{0pt}
$\displaystyle d_{i}$
\end{minipage}};
\draw (451.83,174.5) node   [align=left] {\begin{minipage}[lt]{90.97pt}\setlength\topsep{0pt}
Fuzzing Trajectory
\end{minipage}};
\draw (457.5,206) node   [align=left] {\begin{minipage}[lt]{99.32pt}\setlength\topsep{0pt}
Symbolic Execution \\Trajectory
\end{minipage}};
\draw (436.5,72.5) node   [align=left] {\begin{minipage}[lt]{90.97pt}\setlength\topsep{0pt}
Naive Symbolic\\Execution
\end{minipage}};
\draw (455,122.5) node   [align=left] {\begin{minipage}[lt]{116.55pt}\setlength\topsep{0pt}
Selective Symbolic\\Execution
\end{minipage}};

\end{tikzpicture}
    \caption{Illustrative example of selective symbolic execution to produce input assignment for activating the path $s_i$ to $d_i$. While naive symbolic execution can lead to state space explosion, selective symbolic execution can efficiently solve the constraints. We need to concatenate the respective input assignments of the  fuzzing trajectory with the symbolic execution trajectory.}
    \label{fig:sym_exec}
\end{figure}
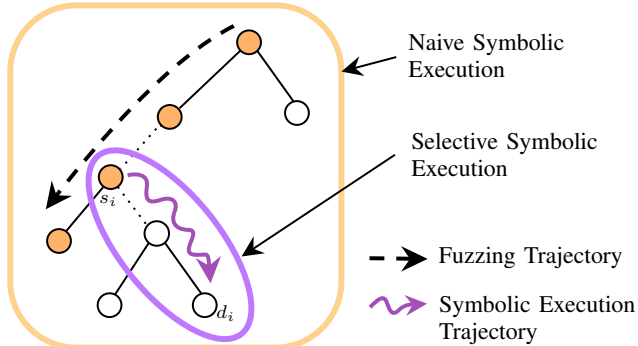

If the constraint is satisfiable, the solver produces a concrete input assignment ($I_c$) that, when executed on the firmware, will traverse the path from $s_i$ to $d_i$, thereby covering the previously unexplored destination node. This concrete input is then added to the fuzzing corpus as illustrated by line 18 in Algorithm~\ref{alg:sysfuss}. Our approach minimizes the cost of symbolic execution by leveraging the output generated during fuzzing to pre-initialize the symbolic state, as illustrated in Figure~\ref{fig:sym_exec}.


The symbolic execution process is repeated for each constraint pair in the priority queue $\mathcal{Q}$ until either all pairs have been processed or a predefined time budget is exhausted, as illustrated by line 12 in Algorithm~\ref{alg:sysfuss}. Each successfully generated test case expands the fuzzing corpus with inputs that can reach previously inaccessible code regions, as in line 18 in Algorithm~\ref{alg:sysfuss}. Once symbolic execution completes, control returns to the fuzzing engine, which resumes mutation-based exploration using the augmented corpus as in line 22 in Algorithm~\ref{alg:sysfuss}. This cycle of fuzzing, plateau detection, symbolic execution, and corpus augmentation continues iteratively until the desired coverage threshold is achieved or the analysis budget is depleted.

\section{Experiments}
\label{sec:experiemnts}

In this section, we present a series of experiments to evaluate our framework (SysFuSS) and demonstrate its effectiveness in addressing the challenges of testing real-world embedded firmware while efficiently uncovering multiple classes of vulnerabilities. We begin by outlining the experimental setup, followed by an assessment of how selective symbolic execution improves analysis of complex firmware binaries. We then compare our implementation with state-of-the-art approaches in terms of branch coverage. Finally, we showcase abilities of our framework in terms of detecting known system-level vulnerabilities and the time required to identify them.


\subsection{Experimental Setup}\label{sec:experimental_setup}

\begin{figure}[htbp]
\centering
\tikzset{every picture/.style={line width=0.75pt}} 

\begin{tikzpicture}[x=0.75pt,y=0.75pt,yscale=-1,xscale=1]

\draw  [fill={rgb, 255:red, 244; green, 217; blue, 255 }  ,fill opacity=1 ] (105,270) -- (415,270) -- (415,375) -- (105,375) -- cycle ;
\draw  [fill={rgb, 255:red, 255; green, 255; blue, 255 }  ,fill opacity=1 ] (115,284.12) .. controls (115,281.84) and (116.84,280) .. (119.12,280) -- (200.88,280) .. controls (203.16,280) and (205,281.84) .. (205,284.12) -- (205,325.88) .. controls (205,328.16) and (203.16,330) .. (200.88,330) -- (119.12,330) .. controls (116.84,330) and (115,328.16) .. (115,325.88) -- cycle ;
\draw  [fill={rgb, 255:red, 222; green, 222; blue, 255 }  ,fill opacity=1 ] (120,290) .. controls (120,287.24) and (122.24,285) .. (125,285) -- (195,285) .. controls (197.76,285) and (200,287.24) .. (200,290) -- (200,305) .. controls (200,307.76) and (197.76,310) .. (195,310) -- (125,310) .. controls (122.24,310) and (120,307.76) .. (120,305) -- cycle ;

\draw  [fill={rgb, 255:red, 255; green, 255; blue, 255 }  ,fill opacity=1 ] (225,290.35) .. controls (225,287.4) and (227.4,285) .. (230.35,285) -- (399.65,285) .. controls (402.6,285) and (405,287.4) .. (405,290.35) -- (405,344.65) .. controls (405,347.6) and (402.6,350) .. (399.65,350) -- (230.35,350) .. controls (227.4,350) and (225,347.6) .. (225,344.65) -- cycle ;
\draw  [fill={rgb, 255:red, 222; green, 222; blue, 255 }  ,fill opacity=1 ] (230,295) .. controls (230,292.24) and (232.24,290) .. (235,290) -- (305,290) .. controls (307.76,290) and (310,292.24) .. (310,295) -- (310,310) .. controls (310,312.76) and (307.76,315) .. (305,315) -- (235,315) .. controls (232.24,315) and (230,312.76) .. (230,310) -- cycle ;
\draw  [fill={rgb, 255:red, 222; green, 222; blue, 255 }  ,fill opacity=1 ] (230,325) .. controls (230,322.24) and (232.24,320) .. (235,320) -- (305,320) .. controls (307.76,320) and (310,322.24) .. (310,325) -- (310,340) .. controls (310,342.76) and (307.76,345) .. (305,345) -- (235,345) .. controls (232.24,345) and (230,342.76) .. (230,340) -- cycle ;
\draw  [fill={rgb, 255:red, 222; green, 222; blue, 255 }  ,fill opacity=1 ] (319.65,295) .. controls (319.65,292.24) and (321.89,290) .. (324.65,290) -- (394.65,290) .. controls (397.41,290) and (399.65,292.24) .. (399.65,295) -- (399.65,310) .. controls (399.65,312.76) and (397.41,315) .. (394.65,315) -- (324.65,315) .. controls (321.89,315) and (319.65,312.76) .. (319.65,310) -- cycle ;
\draw  [fill={rgb, 255:red, 222; green, 222; blue, 255 }  ,fill opacity=1 ] (320,325) .. controls (320,322.24) and (322.24,320) .. (325,320) -- (395,320) .. controls (397.76,320) and (400,322.24) .. (400,325) -- (400,340) .. controls (400,342.76) and (397.76,345) .. (395,345) -- (325,345) .. controls (322.24,345) and (320,342.76) .. (320,340) -- cycle ;
\draw  [fill={rgb, 255:red, 255; green, 178; blue, 105 }  ,fill opacity=1 ] (230,228.25) .. controls (230,226.46) and (231.46,225) .. (233.25,225) -- (286.75,225) .. controls (288.54,225) and (290,226.46) .. (290,228.25) -- (290,246.75) .. controls (290,248.54) and (288.54,250) .. (286.75,250) -- (233.25,250) .. controls (231.46,250) and (230,248.54) .. (230,246.75) -- cycle ;
\draw  [fill={rgb, 255:red, 255; green, 178; blue, 105 }  ,fill opacity=1 ] (105,201.86) .. controls (105,198.07) and (108.07,195) .. (111.86,195) -- (198.14,195) .. controls (201.93,195) and (205,198.07) .. (205,201.86) -- (205,243.14) .. controls (205,246.93) and (201.93,250) .. (198.14,250) -- (111.86,250) .. controls (108.07,250) and (105,246.93) .. (105,243.14) -- cycle ;
\draw  [fill={rgb, 255:red, 255; green, 178; blue, 105 }  ,fill opacity=1 ] (230,198.25) .. controls (230,196.46) and (231.46,195) .. (233.25,195) -- (286.75,195) .. controls (288.54,195) and (290,196.46) .. (290,198.25) -- (290,216.75) .. controls (290,218.54) and (288.54,220) .. (286.75,220) -- (233.25,220) .. controls (231.46,220) and (230,218.54) .. (230,216.75) -- cycle ;
\draw  [fill={rgb, 255:red, 255; green, 178; blue, 105 }  ,fill opacity=1 ] (315,202.4) .. controls (315,198.31) and (318.31,195) .. (322.4,195) -- (402.6,195) .. controls (406.69,195) and (410,198.31) .. (410,202.4) -- (410,242.6) .. controls (410,246.69) and (406.69,250) .. (402.6,250) -- (322.4,250) .. controls (318.31,250) and (315,246.69) .. (315,242.6) -- cycle ;
\draw  [fill={rgb, 255:red, 255; green, 255; blue, 255 }  ,fill opacity=1 ] (115,345) .. controls (115,342.24) and (117.24,340) .. (120,340) -- (200,340) .. controls (202.76,340) and (205,342.24) .. (205,345) -- (205,360) .. controls (205,362.76) and (202.76,365) .. (200,365) -- (120,365) .. controls (117.24,365) and (115,362.76) .. (115,360) -- cycle ;
\draw    (140,250) -- (140,277) ;
\draw [shift={(140,280)}, rotate = 270] [fill={rgb, 255:red, 0; green, 0; blue, 0 }  ][line width=0.08]  [draw opacity=0] (8.04,-3.86) -- (0,0) -- (8.04,3.86) -- (5.34,0) -- cycle    ;
\draw    (255,270) -- (255,253) ;
\draw [shift={(255,250)}, rotate = 90] [fill={rgb, 255:red, 0; green, 0; blue, 0 }  ][line width=0.08]  [draw opacity=0] (8.04,-3.86) -- (0,0) -- (8.04,3.86) -- (5.34,0) -- cycle    ;
\draw    (270,250) -- (270,267) ;
\draw [shift={(270,270)}, rotate = 270] [fill={rgb, 255:red, 0; green, 0; blue, 0 }  ][line width=0.08]  [draw opacity=0] (8.04,-3.86) -- (0,0) -- (8.04,3.86) -- (5.34,0) -- cycle    ;
\draw    (290,240) -- (312,240) ;
\draw [shift={(315,240)}, rotate = 180] [fill={rgb, 255:red, 0; green, 0; blue, 0 }  ][line width=0.08]  [draw opacity=0] (8.04,-3.86) -- (0,0) -- (8.04,3.86) -- (5.34,0) -- cycle    ;
\draw    (315,205) -- (293,205) ;
\draw [shift={(290,205)}, rotate = 360] [fill={rgb, 255:red, 0; green, 0; blue, 0 }  ][line width=0.08]  [draw opacity=0] (8.04,-3.86) -- (0,0) -- (8.04,3.86) -- (5.34,0) -- cycle    ;
\draw    (230,205) -- (208,205) ;
\draw [shift={(205,205)}, rotate = 360] [fill={rgb, 255:red, 0; green, 0; blue, 0 }  ][line width=0.08]  [draw opacity=0] (8.04,-3.86) -- (0,0) -- (8.04,3.86) -- (5.34,0) -- cycle    ;
\draw    (205,215) -- (227,215) ;
\draw [shift={(230,215)}, rotate = 180] [fill={rgb, 255:red, 0; green, 0; blue, 0 }  ][line width=0.08]  [draw opacity=0] (8.04,-3.86) -- (0,0) -- (8.04,3.86) -- (5.34,0) -- cycle    ;
\draw    (175,250) -- (175,260) -- (215,260) -- (215,320) -- (222,320) ;
\draw [shift={(225,320)}, rotate = 180] [fill={rgb, 255:red, 0; green, 0; blue, 0 }  ][line width=0.08]  [draw opacity=0] (8.04,-3.86) -- (0,0) -- (8.04,3.86) -- (5.34,0) -- cycle    ;

\draw (260,237.5) node  [font=\scriptsize] [align=left] {\begin{minipage}[lt]{40.8pt}\setlength\topsep{0pt}
\begin{center}
\textbf{Coverage}\\\textbf{Analyzer}
\end{center}

\end{minipage}};
\draw (155,222.5) node  [font=\scriptsize] [align=left] {\begin{minipage}[lt]{68pt}\setlength\topsep{0pt}
\begin{center}
\textbf{Fuzzing Engine}\\(LibAFL)
\end{center}

\end{minipage}};
\draw (267.5,367.5) node  [font=\scriptsize] [align=left] {\begin{minipage}[lt]{221pt}\setlength\topsep{0pt}
\begin{center}
\textbf{QEMU Emulator}
\end{center}

\end{minipage}};
\draw (260,207.5) node  [font=\scriptsize] [align=left] {\begin{minipage}[lt]{40.8pt}\setlength\topsep{0pt}
\begin{center}
\textbf{Corpus}
\end{center}

\end{minipage}};
\draw (362.5,222.5) node  [font=\scriptsize] [align=left] {\begin{minipage}[lt]{64.6pt}\setlength\topsep{0pt}
\begin{center}
\textbf{Symbolic Execution Engine}\\(angr)
\end{center}

\end{minipage}};
\draw (160,297.5) node  [font=\scriptsize] [align=left] {\begin{minipage}[lt]{54.4pt}\setlength\topsep{0pt}
\begin{center}
Code Hooks
\end{center}

\end{minipage}};
\draw (160,320) node  [font=\scriptsize] [align=left] {\begin{minipage}[lt]{61.2pt}\setlength\topsep{0pt}
\begin{center}
\textbf{TCG Guest Code}
\end{center}

\end{minipage}};
\draw (160,352.5) node  [font=\scriptsize] [align=left] {\begin{minipage}[lt]{61.2pt}\setlength\topsep{0pt}
\begin{center}
\textbf{vCPUs}
\end{center}

\end{minipage}};
\draw (270,302.5) node  [font=\scriptsize] [align=left] {\begin{minipage}[lt]{54.4pt}\setlength\topsep{0pt}
\begin{center}
Memory Hooks
\end{center}

\end{minipage}};
\draw (270,332.5) node  [font=\scriptsize] [align=left] {\begin{minipage}[lt]{54.4pt}\setlength\topsep{0pt}
\begin{center}
Syscall Hooks
\end{center}

\end{minipage}};
\draw (359.65,302.5) node  [font=\scriptsize] [align=left] {\begin{minipage}[lt]{54.4pt}\setlength\topsep{0pt}
\begin{center}
Crash Handlers
\end{center}

\end{minipage}};
\draw (360,332.5) node  [font=\scriptsize] [align=left] {\begin{minipage}[lt]{54.4pt}\setlength\topsep{0pt}
\begin{center}
Thread Hooks
\end{center}

\end{minipage}};

\end{tikzpicture}
\caption{Architecture of the proposed LibAFL-QEMU integration showing symbolic execution and in-emulator instrumentation for coverage and memory safety tracking.}
\label{fig:qemu_libafl_arch}
\end{figure}
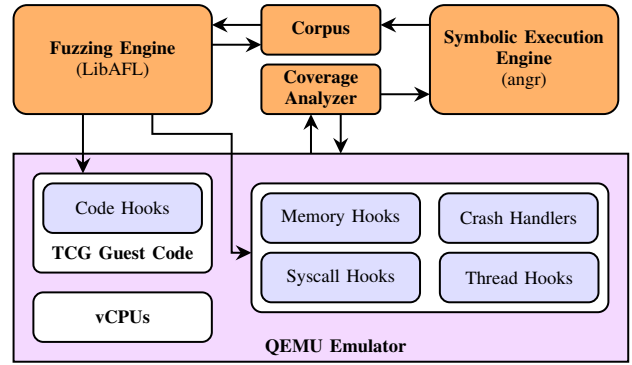

We implemented our framework on top of \textit{LibAFL}~\cite{malmain2024libafl}, a modular Rust-based fuzzing infrastructure that supports full-system emulation via \textit{QEMU}~\cite{bellard2005qemu}. To meet our design goals, we extended LibAFL with RISC-V architecture support, memory safety instrumentation, and symbolic execution capabilities, as illustrated in Figure~\ref{fig:qemu_libafl_arch}.

\vspace{0.1in}
\noindent\textbf{Architecture Support:} 
LibAFL’s native backend supports common ISAs such as x86 and ARM. We extended both LibAFL and its QEMU integration layer to include \textit{RISC-V} by adding architecture-aware register definitions, hooks for program counter tracking, and system call handling. Corresponding updates to the QEMU bridge expose instrumentation points specific to RISC-V through the Tiny Code Generator (TCG), enabling coverage collection and execution control during fuzzing.

\vspace{0.1in}
\noindent\textbf{Memory Safety Instrumentation:} 
We instrumented QEMU’s \texttt{SoftMMU} subsystem to monitor heap memory operations using lightweight hooks inserted before and after each access. A shadow memory mirrors the guest heap, annotating bytes as \emph{Unaddressable}, \emph{Uninitialized}, \emph{Defined}, or \emph{Readable} as discussed in Section~\ref{sec:memvul_detect}. Access violations, such as use-after-free or uninitialized reads, are detected in real time and reported to the LibAFL harness via Foreign Function Interface (FFI), with minimal performance overhead to the entire framework.

\begin{table}[htb]
\centering
\vspace{0.1in}
\caption{Firmware benchmarks used for evaluation}
\label{tab:emd_firm_bench}
\begin{tabular}{@{}llcr@{}}
\toprule
\textbf{Firmware} & \multicolumn{1}{l}{\textbf{Type}} & \textbf{Architecture} & \textbf{Size(KB)}\\ \midrule
OpenSSL & Cryptographic library & x86 & 2,450\\ \midrule
WolfBoot & Secure bootloader & RISC-V & 180\\ \midrule
WolfMQTT & MQTT client library & x86 & 95\\ \midrule
HTSlib & Seq. data format library & Arm & 1,230\\ \midrule
MXML & Mini XML parser & Arm & 65\\ \midrule
libIEC & Substation automation sys. & Arm & 890\\ \bottomrule
\end{tabular}
\end{table}

\begin{figure*}[!htp]
    \centering
    \resizebox{\textwidth}{!}{\input{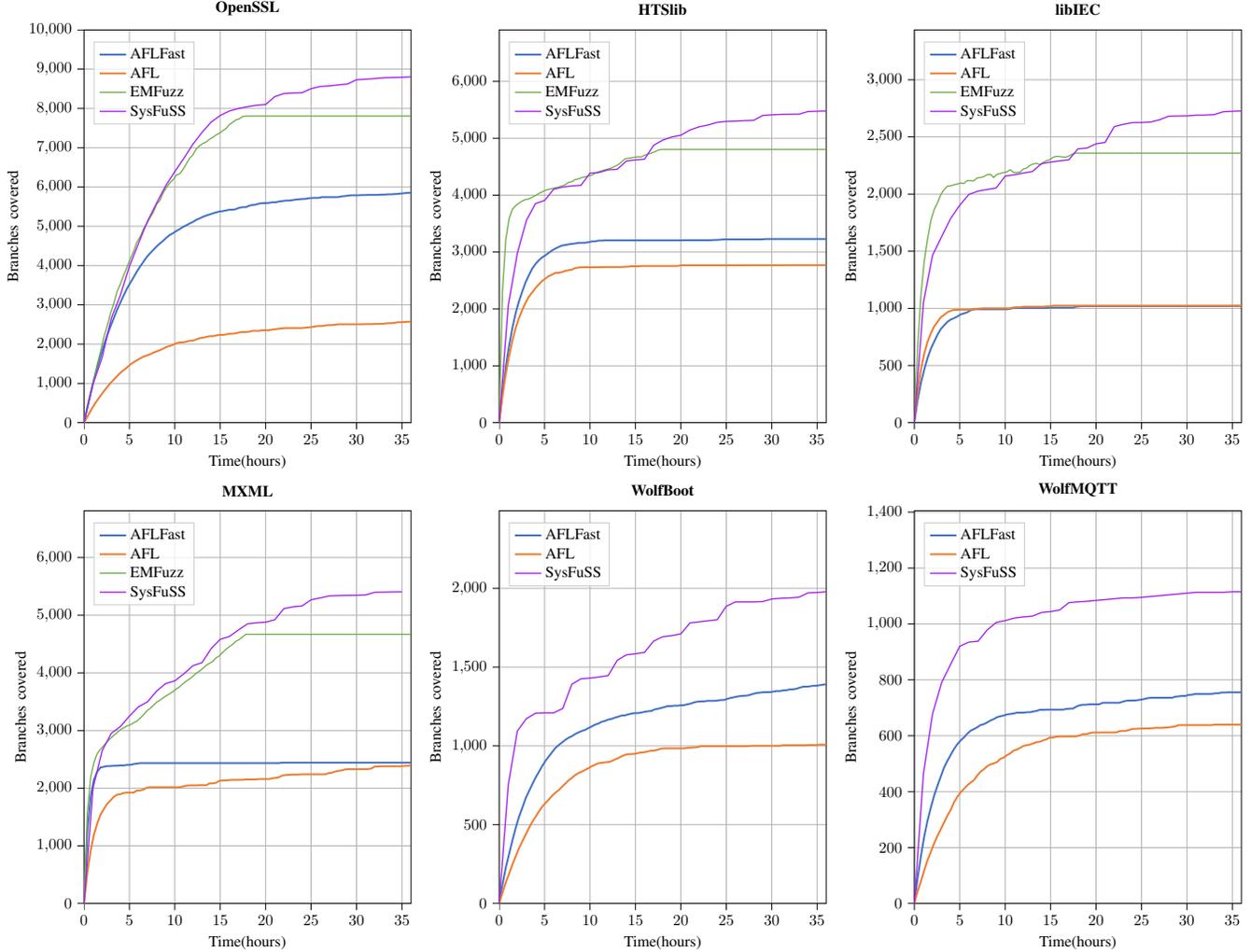}}
    \caption{Branch coverage comparison across six firmware benchmarks. SysFuSS consistently outperforms baseline tools, achieving 1.3--1.9× improvement over AFL, AFLFast, and EM-Fuzz. Note that we have obtained the EM-Fuzz results for OpenSSL, HTSlib, libIEC, and MXML from~\cite{gao2020fuzz}. Since the authors did not provide the source code, we could not generate EM-Fuzz results for WolfBoot and WolfMQTT.}
    \label{fig:coverage_results}
\end{figure*}
\vspace{0.1in}
\noindent\textbf{Symbolic Execution: }
For guided input generation and binary analysis, we integrated \textit{angr}~\cite{wang2017angr}, which lifts machine code to an intermediate representation and performs path constraint solving via the \textit{Z3} SMT solver~\cite{de2008z3}. Each symbolic path produces constraints $\phi_{s_i \rightarrow d_i}$, solved to generate concrete inputs that drive the fuzzer toward unexplored states. All generated inputs are validated through concrete QEMU execution to ensure soundness.

\begin{table*}[!ht]
\centering
\caption{Vulnerabilities present in each firmware (2016-2025) and detection results. SysFuSS can detect {118} vulnerabilities out of 223 known CVEs. It may be possible to detect the remaining CVEs by SysFuSS if we extend the fuzzing time.}
\label{tab:vulnerability_categories}

\begin{tabular}{@{}|l|rr|rr|rr|rr|rr|rr|@{}}
\toprule
\multirow{2}{*}{\textbf{Category}} & \multicolumn{2}{c|}{\textbf{OpenSSL}} & \multicolumn{2}{c|}{\textbf{HTSlib}} & \multicolumn{2}{c|}{\textbf{libIEC}} & \multicolumn{2}{c|}{\textbf{Mini-XML}} & \multicolumn{2}{c|}{\textbf{WolfBoot}} & \multicolumn{2}{c|}{\textbf{WolfMQTT}} \\ \cmidrule(l){2-13} 
 & \multicolumn{1}{c|}{CVE} & \multicolumn{1}{c|}{SysFuSS} & \multicolumn{1}{c|}{CVE} & \multicolumn{1}{c|}{SysFuSS} & \multicolumn{1}{c|}{CVE} & \multicolumn{1}{c|}{SysFuSS} & \multicolumn{1}{c|}{CVE} & \multicolumn{1}{c|}{SysFuSS} & \multicolumn{1}{c|}{CVE} & \multicolumn{1}{c|}{SysFuSS} & \multicolumn{1}{c|}{CVE} & \multicolumn{1}{c|}{SysFuSS} \\ \midrule
Memory Safety Issues & \multicolumn{1}{r|}{40} & 25 & \multicolumn{1}{r|}{6} & 5 & \multicolumn{1}{r|}{21} & 15 & \multicolumn{1}{r|}{6} & 4 & \multicolumn{1}{r|}{13} & 10 & \multicolumn{1}{r|}{8} & 5 \\ \midrule
Protocol Design & \multicolumn{1}{r|}{15} & 10 & \multicolumn{1}{r|}{1} & 1 & \multicolumn{1}{r|}{2} & 1 & \multicolumn{1}{r|}{-} & - & \multicolumn{1}{r|}{5} & 1 & \multicolumn{1}{r|}{-} & - \\ \midrule
Denial of Services & \multicolumn{1}{r|}{20} & 10 & \multicolumn{1}{r|}{1} & - & \multicolumn{1}{r|}{5} & 3 & \multicolumn{1}{r|}{2} & 1 & \multicolumn{1}{r|}{8} & 3 & \multicolumn{1}{r|}{-} & - \\ \midrule
Side-Channel Attacks & \multicolumn{1}{r|}{12} & 3 & \multicolumn{1}{r|}{-} & - & \multicolumn{1}{r|}{-} & - & \multicolumn{1}{r|}{-} & - & \multicolumn{1}{r|}{18} & 2 & \multicolumn{1}{r|}{-} & - \\ \midrule
Cryptographic Issues & \multicolumn{1}{r|}{12} & 5 & \multicolumn{1}{r|}{-} & - & \multicolumn{1}{r|}{-} & - & \multicolumn{1}{r|}{-} & - & \multicolumn{1}{r|}{6} & 2 & \multicolumn{1}{r|}{-} & - \\ \midrule
Certificate Verification & \multicolumn{1}{r|}{15} & 7 & \multicolumn{1}{r|}{-} & - & \multicolumn{1}{r|}{-} & - & \multicolumn{1}{r|}{-} & - & \multicolumn{1}{r|}{9} & 5 & \multicolumn{1}{r|}{-} & - \\ \midrule
\textbf{Total} & \multicolumn{1}{r|}{\textbf{114}} & \textbf{60} & \multicolumn{1}{r|}{\textbf{8}} & \textbf{6} & \multicolumn{1}{r|}{\textbf{28}} & \textbf{19} & \multicolumn{1}{r|}{\textbf{6}} & \textbf{5} & \multicolumn{1}{r|}{\textbf{59}} & \textbf{23} & \multicolumn{1}{r|}{\textbf{8}} & \textbf{5} \\ \bottomrule
\end{tabular}
\end{table*}

\vspace{0.1in}
\noindent\textbf{Benchmarks: } In order to evaluate and compare the results of SysFuSS, we have selected six real-world system-level firmware implementations spanning cryptography (OpenSSL), bootloaders (WolfBoot), communication protocols (WolfMQTT), and data processing libraries (HTSlib, MXML, libIEC). Table~\ref{tab:emd_firm_bench} presents the benchmark firmware used in our evaluation, including their sizes and primary functionalities.

\vspace{0.3in}
\noindent\textbf{Evaluation Environment: } All experiments were conducted on a 64-bit machine equipped with a 40-core Intel Xeon E5-2698 v4 CPU running at 2.20GHz, 512GB of RAM, and Ubuntu 22.04.5 LTS as the host operating system. For each benchmark, we executed the fuzzing with symbolic execution for 36 hours. To account for the inherent randomness in coverage-guided fuzzing, we repeat each experiment five times using identical initial seed inputs. We compare SysFuSS against three state-of-the-art firmware fuzzing tools: AFL~\cite{fioraldi2020afl++}, AFLFast~\cite{bohme2016coverage}, and EM-Fuzz~\cite{gao2020fuzz}. AFL represents the baseline coverage-guided fuzzer, AFLFast improves upon AFL with optimized power scheduling, and EM-Fuzz targets embedded firmware with memory-aware fuzzing.

\subsection{Impact of Selective Symbolic Execution}\label{sec:rq1}

\begin{figure}[ht]
\centering
\footnotesize
\begin{tikzpicture}

\definecolor{blueviolet16032240}{RGB}{160,32,240}
\definecolor{darkgray176}{RGB}{176,176,176}
\definecolor{darkorange25512714}{RGB}{255,127,14}
\definecolor{lightgray204}{RGB}{204,204,204}

\begin{axis}[width=8.2cm,height=6cm,
legend cell align={left},
legend style={
  fill opacity=0.8,
  draw opacity=1,
  text opacity=1,
  at={(0.03,0.97)},
  anchor=north west,
  draw=lightgray204
},
minor xtick={},
minor ytick={},
tick align=outside,
tick pos=left,
x grid style={darkgray176},
xlabel={Time (hours)},
xmajorgrids,
xmin=0, xmax=36,
xtick style={color=black},
xtick={0,4,8,12,16,20,24,28,32,36},
y grid style={darkgray176},
ylabel={Branches Covered},
ymajorgrids,
ymin=0, ymax=2500,
ytick style={color=black},
ytick={0,500,1000,1500,2000,2500}
]
\addplot [very thick, blueviolet16032240]
table {%
0 0
1 752.1
2 1094.8
3 1173
4 1207.5
5 1209.8
6 1209.8
7 1237.4
8 1391.5
9 1426
10 1430.6
11 1437.5
12 1446.7
13 1543.3
14 1577.8
15 1584.7
16 1593.9
17 1665.2
18 1692.8
19 1699.7
20 1711.2
21 1780.2
22 1787.1
23 1794
24 1800.9
25 1886
26 1913.6
27 1913.6
28 1913.6
29 1915.9
30 1932
31 1936.6
32 1938.9
33 1943.5
34 1971.1
35 1973.4
36 1978
};
\addlegendentry{SysFuSS with Symbolic Execution}
\addplot [very thick, darkorange25512714]
table {%
0 0
1 768.2
2 1099.4
3 1150
4 1196
5 1198.3
6 1207.5
7 1242
8 1267.3
9 1281.1
10 1288
11 1288
12 1301.8
13 1311
14 1331.7
15 1334
16 1334
17 1334
18 1336.3
19 1336.3
20 1352.4
21 1352.4
22 1368.5
23 1368.5
24 1370.8
25 1373.1
26 1377.7
27 1380
28 1391.5
29 1398.4
30 1409.9
31 1421.4
32 1442.1
33 1446.7
34 1449
35 1451.3
36 1451.3
};
\addlegendentry{SysFuSS without Symbolic Execution}
\end{axis}

\end{tikzpicture}
\caption{Branch coverage for WolfBoot using SysFuSS (with and without symbolic execution).}
\label{fig:comparison}
\end{figure}
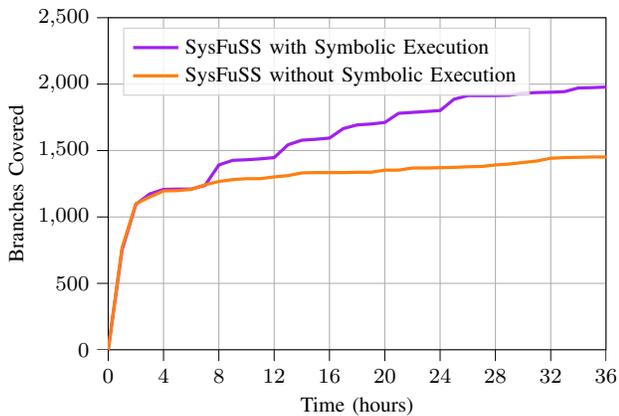

In order to assess the impact of integrating selective symbolic execution with the coverage plateau detection, we compared SysFuSS with symbolic execution against SysFuSS without symbolic execution on WolfBoot benchmark. Figure~\ref{fig:comparison} shows branch coverage over 36 hours for the WolfBoot benchmark. SysFuSS without symbolic execution plateaued at roughly 60\% coverage after 8 hours (covered 1500 out of 2500 branches), as random mutations failed to satisfy complex path constraints.
In contrast, SysFuSS with symbolic execution exhibited periodic coverage jumps triggered by the plateau detection mechanism (Section~\ref{sec:plateau_detection}), leading to 80\% branch coverage (covering 2000 out of 2500 branches). Each invocation of symbolic execution generated targeted inputs that unlocked previously unreachable regions, allowing subsequent fuzzing to expand coverage rapidly. We have observed the same trend for the remaining firmware benchmarks.

\begin{table*}[!ht]
\centering
\caption{Time (in hours) to detect vulnerabilities within 36 hours. The improvement is over the best detection time (shown in bold). SysFuSS significantly outperforms existing fuzzers in terms of vulnerability detection time.}
\label{tab:time}
\small
\begin{tabular}{@{}|cll|c|c|c|c|c|@{}}
\toprule
\multicolumn{1}{|c|}{\textbf{Program}} & \multicolumn{1}{l|}{\textbf{Identifier}} & \textbf{Type} & \textbf{AFL} & \textbf{AFLFast} & \textbf{EM-Fuzz} & \textbf{SysFuSS} & \textbf{Improvement} (over best) \\ \midrule
\multicolumn{1}{|c|}{OpenSSL} & \multicolumn{1}{l|}{\begin{tabular}[c]{@{}l@{}}CVE-2016-2108\end{tabular}} & \begin{tabular}[c]{@{}l@{}}buffer underflow\end{tabular} & \begin{tabular}[c]{@{}c@{}}\textbf{5}\end{tabular} & \begin{tabular}[c]{@{}c@{}}6\end{tabular} & \begin{tabular}[c]{@{}c@{}}13\end{tabular} & \begin{tabular}[c]{@{}c@{}}3\end{tabular} & \begin{tabular}[c]{@{}c@{}} 1.67x\end{tabular} \\ \midrule
\multicolumn{1}{|c|}{HTSlib} & \multicolumn{1}{l|}{\begin{tabular}[c]{@{}l@{}}CVE-2018-13843\\ CVE-2018-13844\\ CVE-2018-13845\end{tabular}} & \begin{tabular}[c]{@{}l@{}}memory leak\\ memory leak\\ buffer over-read\end{tabular} & \begin{tabular}[c]{@{}c@{}}-\\ -\\ -\end{tabular} & \begin{tabular}[c]{@{}c@{}}-\\ -\\ -\end{tabular} & \begin{tabular}[c]{@{}c@{}}\textbf{7}\\ \textbf{8}\\ \textbf{5}\end{tabular} & \begin{tabular}[c]{@{}c@{}}4\\ 5\\ 3\end{tabular} & \begin{tabular}[c]{@{}c@{}}1.75x\\ 1.60x\\ 1.67x\end{tabular} \\ \midrule
\multicolumn{1}{|c|}{IEC61850} & \multicolumn{1}{l|}{\begin{tabular}[c]{@{}l@{}}CVE-2018-18834\\ CVE-2018-18937\\ CVE-2018-19093\\ CVE-2018-19185\\ CVE-2018-19121\\ CVE-2018-19122\end{tabular}} & \begin{tabular}[c]{@{}l@{}}buffer overflow\\ NULL ptr dereference\\ segmentation fault\\ buffer overflow\\ segmentation fault\\ NULL ptr dereference\end{tabular} & \begin{tabular}[c]{@{}c@{}}10\\ 15\\ 22\\ -\\ -\\ -\end{tabular} & \begin{tabular}[c]{@{}c@{}}9\\ 14\\ 17\\ 21\\ -\\ -\end{tabular} & \begin{tabular}[c]{@{}c@{}}\textbf{3}\\ \textbf{7}\\ \textbf{11}\\ \textbf{2}\\ \textbf{8}\\ \textbf{13}\end{tabular} & \begin{tabular}[c]{@{}c@{}}2\\ 4\\  4\\ 2\\ 7\\ 9\end{tabular} & \begin{tabular}[c]{@{}c@{}}1.50x\\ 1.75x\\ 2.75x\\ 1.00x\\ 1.14x\\ 1.44x\end{tabular} \\ \midrule
\multicolumn{1}{|c|}{MXML} & \multicolumn{1}{l|}{\begin{tabular}[c]{@{}l@{}}CVE-2018-19764\\ CVE-2018-20004\\ CVE-2018-20005\end{tabular}} & \begin{tabular}[c]{@{}l@{}}stack overflow\\ use-after-free\\ memory leak\end{tabular} & \begin{tabular}[c]{@{}c@{}}19\\ -\\ -\end{tabular} & \begin{tabular}[c]{@{}c@{}} 7\\ -\\ -\end{tabular} & \begin{tabular}[c]{@{}c@{}} \textbf{3}\\ \textbf{6}\\ \textbf{5}\end{tabular} & \begin{tabular}[c]{@{}c@{}} 3\\ 4\\ 3\end{tabular} & \begin{tabular}[c]{@{}c@{}}1.00x\\ 1.50x\\ 1.67x\end{tabular} \\ \midrule
\multicolumn{1}{|c|}{WolfBoot} & \multicolumn{1}{l|}{\begin{tabular}[c]{@{}l@{}}CVE-2024-5991\end{tabular}} & \begin{tabular}[c]{@{}l@{}}NULL ptr dereference\end{tabular} & \begin{tabular}[c]{@{}c@{}}11\end{tabular} & \begin{tabular}[c]{@{}c@{}}\textbf{10}\end{tabular} & \begin{tabular}[c]{@{}c@{}}-\end{tabular} & \begin{tabular}[c]{@{}c@{}}3\end{tabular} & \begin{tabular}[c]{@{}c@{}}3.33x\end{tabular} \\ \midrule
\multicolumn{3}{|c|}{\textbf{Average}} & \textbf{13.8} & \textbf{11.7} & \textbf{7.0} & \textbf{4.0} & \textbf{1.75x} \\ \bottomrule
\end{tabular}
\end{table*}

\subsection{Comparison of Branch Coverage}\label{sec:rq2}

To evaluate the overall effectiveness of our framework, we compared SysFuSS with state-of-the-art fuzzing tools  (\textit{EM-Fuzz}~\cite{gao2020fuzz}, \textit{AFLFast}~\cite{bohme2016coverage}, and \textit{AFL}~\cite{afl})  across six real-world firmware benchmarks. Figure~\ref{fig:coverage_results} presents the branch coverage achieved by each tool over a 36-hour fuzzing period. The objective of this experiment is to assess how the integration of selective symbolic execution and system-level emulation influences coverage progression compared to purely mutation-based fuzzers.
As illustrated by Figure~\ref{fig:coverage_results}, our approach consistently provides highest branch coverage across all benchmarks. For instance, in the OpenSSL benchmark, it reaches approximately 9,000 branches; surpassing EM-Fuzz, AFLFast, and AFL by 12\%, 48\%, and 300\%, respectively. It is evident that existing fuzzers exhibit clear coverage plateaus, where further mutations yield diminishing returns. SysFuSS effectively breaks through these plateaus by invoking symbolic execution when coverage hits a wall. For example, in WolfBoot, SysFuSS extends coverage close to 2,000 branches, while AFL and AFLFast can cover only 1,400 branches. 
Overall, these results demonstrate that SysFuSS is able to provide higher branch coverage than conventional fuzzers due to effective utilization of selective symbolic execution.

\subsection{Detection of Known Vulnerabilities}\label{sec:rq3}

Table~\ref{tab:vulnerability_categories} provides a comprehensive breakdown of vulnerability categories present in each firmware target and demonstrates SysFuSS's detection capabilities across diverse vulnerability types spanning from 2016 to 2025 from the selected benchmarks in Table~\ref{tab:emd_firm_bench}. Across all targets, SysFuSS successfully detected 118 vulnerabilities out of 223 known CVEs, representing a detection rate of approximately 47\%. The results show varying detection rates across different firmware, with OpenSSL yielding 60 detections out of 114 CVEs, libIEC achieving 19 out of 28, and WolfBoot reaching 23 out of 59. Memory safety issues constitute the most prevalent vulnerability category, with SysFuSS detecting 64 such vulnerabilities across all firmware targets. Beyond memory safety, the framework demonstrates effectiveness in identifying protocol design flaws (13 detections), denial of service vulnerabilities (17 detections), and various security-critical issues including side-channel attacks, cryptographic weaknesses, and certificate verification problems. Notably, certain firmware like HTSlib and WolfMQTT show higher detection rates relative to their total CVE counts, while OpenSSL's lower detection rate reflects the inherent complexity and vast attack surface of cryptographic libraries.

Table~\ref{tab:vulnerability_comparison} compares the number of known vulnerabilities (CVEs) in the six benchmarks that can be detected by different fuzzers. We have obtained the numbers for the existing methods from their respective publications. The exact vulnerabilities they can cover (and their detection time) is also outlined in Table~\ref{tab:time}. The total number of vulnerabilities detected by our approach is obtained from Table~\ref{tab:vulnerability_categories}. SysFuSS can detect significantly more vulnerabilities compared to existing methods. Specifically, while SysFuss can detect 118 vulnerabilities, AFL, AFLFast and EM-Fuzz can detect only 6, 7, and 13 vulnerabilities, respectively.

\begin{table}[htp]
    \centering
    \caption{Number of known vulnerabilities (CVEs) in the six benchmarks detected by various approaches}
    \label{tab:vulnerability_comparison}
    \begin{tabular}{|c|c|c|c|}
    \hline
        \textbf{AFL}~\cite{afl} & \textbf{AFLFast}~\cite{bohme2016coverage} & \textbf{EM-Fuzz}~\cite{gao2020fuzz} & \textbf{SysFuSS} \\ \hline
        6 & 7 & 13 & 118 \\ \hline
    \end{tabular}
\end{table}

\subsection{Comparison of Vulnerability Detection Time}\label{sec:rq4}

Table~\ref{tab:time} summarizes the detection time for various vulnerabilities across six firmware benchmarks. The first column provides the benchmarks. The second column lists the vulnerabilities that have been detected by any of the existing fuzzers. The third column shows the type of attack. The next four columns show the detection time (in hours) by various fuzzers. The last column shows the improvement in detection time by SysFuSS compared to the best known result (marked in bold). The results highlight two important aspects. First, SysFuSS can significantly outperform in detection time (up to 3.33 times, 1.75 times on average) compared to the state-of-the-art approaches. Next, SysFuSS can detect the vulnerabilities that can be detected by existing methods. In fact, as discussed in Section~\ref{sec:rq3}, SysFuSS can detect significantly more vulnerabilities (118 versus 13) compared to state-of-the-art approaches.

\section{Conclusion}
\label{sec:conclusion}

Firmware security remains a pressing challenge as the rapid expansion of IoT and embedded systems continues to widen the attack surface across critical infrastructure and consumer technologies. Traditional fuzzing methods, while effective for user-space applications, often fail to achieve deep coverage in firmware due to hardware dependencies and complex path constraints.

This paper presented SysFuSS, a hybrid firmware verification framework, that integrates system-level emulation with selective symbolic execution. Our framework introduces three core advancements: (1) full-system emulation via QEMU for comprehensive firmware analysis beyond user-space analysis, (2) selective symbolic execution that dynamically alternates between fuzzing and symbolic execution to handle the vast state space introduced by the system-level firmware, and (3) shadow memory instrumentation for precise, real-time detection of memory corruption vulnerabilities.

Extensive evaluation using six real-world firmware benchmarks demonstrated that SysFuSS can significantly improve branch coverage (12\% on average) over state-of-the-art fuzzers. SysFuSS can also detect significantly more vulnerabilities (118 versus 13) than the state-of-the-art. The results also highlight that SysFuSS can detect the vulnerabilities significantly faster (up to 3.33$\times$) compared to the best known results for each vulnerability. These results confirm that SysFuSS effectively bridges the gap between scalability and depth in firmware testing, enabling efficient exploration of embedded firmware for security analysis.

\ifCLASSOPTIONcompsoc


\bibliographystyle{IEEEtran}
\bibliography{bibliography}

\end{document}